\documentclass[aps,prb,twocolumn,superscriptaddress,floatfix,letter]{revtex4-1}
\usepackage[bookmarks=true,colorlinks,linkcolor=OrangeRed,urlcolor=NavyBlue,citecolor=RoyalBlue]{hyperref}
\usepackage{amsmath,amssymb,color,comment,physics}
\usepackage[makeroom]{cancel}
\usepackage[caption=false]{subfig}
\usepackage{mathrsfs}
\usepackage{graphicx}
\usepackage{subfig}
\usepackage[countmax]{subfloat}
\usepackage[english]{babel}
\usepackage[dvipsnames]{xcolor}
\usepackage{mathtools}

\usepackage{ulem}

\usepackage{braket}

\definecolor{mypurple}{rgb}{0.49,0.18,0.56}
\definecolor{mygold}{rgb}{0.93,0.49,0.13}
\definecolor{mygreen}{rgb}{0,0.5,0}
\definecolor{myblue}{rgb}{0,0,0.75}
\definecolor{mymagenta}{cmyk}{0,1,0,0.12}
\definecolor{mygray}{rgb}{0.5,0.5,0.5}

\usepackage{umoline}

\definecolor{mypink1}{rgb}{0.858, 0.188, 0.478}

\begin{document}

\title{Reliability of lattice gauge theories in the  thermodynamic limit}
\author{Maarten Van Damme}
\affiliation{Department of Physics and Astronomy, University of Ghent, Krijgslaan 281, 9000 Gent, Belgium}

\author{Haifeng Lang}
\affiliation{INO-CNR BEC Center and Department of Physics, University of Trento, Via Sommarive 14, I-38123 Trento, Italy}
\affiliation{Theoretical Chemistry, Institute of Physical Chemistry,
Heidelberg University, Im Neuenheimer Feld 229, 69120 Heidelberg, Germany}

\author{Philipp Hauke}
\affiliation{INO-CNR BEC Center and Department of Physics, University of Trento, Via Sommarive 14, I-38123 Trento, Italy}

\author{Jad C.~Halimeh}
\affiliation{INO-CNR BEC Center and Department of Physics, University of Trento, Via Sommarive 14, I-38123 Trento, Italy}

\begin{abstract}
	Although gauge invariance is a postulate in fundamental theories of nature such as quantum electrodynamics, in quantum-simulation implementations of gauge theories it is compromised by experimental imperfections. In a recent work [Halimeh and Hauke, \href{https://journals.aps.org/prl/abstract/10.1103/PhysRevLett.125.030503}{Phys.~Rev.~Lett.~\textbf{125}, 030503 (2020)}], it has been shown in finite-size spin-$1/2$ quantum link lattice gauge theories that upon introducing an energy-penalty term of sufficiently large strength $V$, unitary gauge-breaking errors at strength $\lambda$ are suppressed $\propto\lambda^2/V^2$ up to all accessible evolution times. Here, we show numerically that this result extends to quantum link models in the thermodynamic limit and with larger spin-$S$. As we show analytically, the dynamics at short times is described by an \textit{adjusted} gauge theory up to a timescale that is at earliest $\tau_\text{adj}\propto\sqrt{V/V_0^3}$, with $V_0$ an energy factor. Moreover, our analytics predicts that a renormalized gauge theory dominates at intermediate times up to a timescale $\tau_\text{ren}\propto\exp(V/V_0)/V_0$. In both emergent gauge theories, $V$ is volume-independent and scales at worst $\sim S^2$. Furthermore, we numerically demonstrate that robust gauge invariance is also retained through a single-body gauge-protection term, which is experimentally straightforward to implement in ultracold-atom setups and NISQ devices.
\end{abstract}
\date{\today}
\maketitle
\tableofcontents
\section{Introduction}
Recent years have witnessed impressive progress in the level of control and precision achieved in synthetic quantum matter.\cite{Bloch2008,Lewenstein_book,Blatt_review,Hauke2012}  In addition to allowing for the exploration of exotic phenomena such as many-body localization,\cite{Schreiber2015,Choi2016,Smith2016} the Kibble-Zurek mechanism,\cite{Xu2014,Anquez2016,Clark2016,Cui2016,Keesling2019} dynamical phase transitions,\cite{Jurcevic2017,Zhang2017dpt,Flaeschner2018} prethermalization,\cite{Gring2012,Langen2015,Neyenhuis2017} and many-body dephasing,\cite{Kaplan2020} this technological advancement has facilitated the realization of complex multi-species systems such as lattice gauge theories. \cite{Martinez2016,Bernien2017,Dai2017,Klco2018,Kokail2019,Schweizer2019,Goerg2019,Mil2020,Brower2020,Klco2020,Yang2020} Not only can this ability enable a possible foray into questions of high-energy physics in inexpensive low-energy tabletop quantum simulators,\cite{Wiese_review,Zohar2015,Dalmonte2016,MariCarmen2019} it also sets the grounds for a standard experimental benchmark for the latter. Indeed, even though in nature gauge invariance is a postulate, such as Gauss's law in quantum electrodynamics (QED), it is not guaranteed in generic ultracold-atom implementations of lattice gauge theories (excepting for realizations that use manifestly gauge-invariant mappings to reduce the number of degrees of freedom\cite{Zohar2015,Muschik2017,Surace2020}). This potential detriment is due to experimental errors avoidable only by way of unrealistic fine-tuning in the experimental parameters. Amidst the current intense drive in both academia and industry to build reliable quantum simulators, gauge-theory implementations therefore promise to serve as a measure of experimental control and precision regarding how robustly gauge invariance can be enforced.

\begin{figure}[htp]
	\centering
	\includegraphics[width=.48\textwidth]{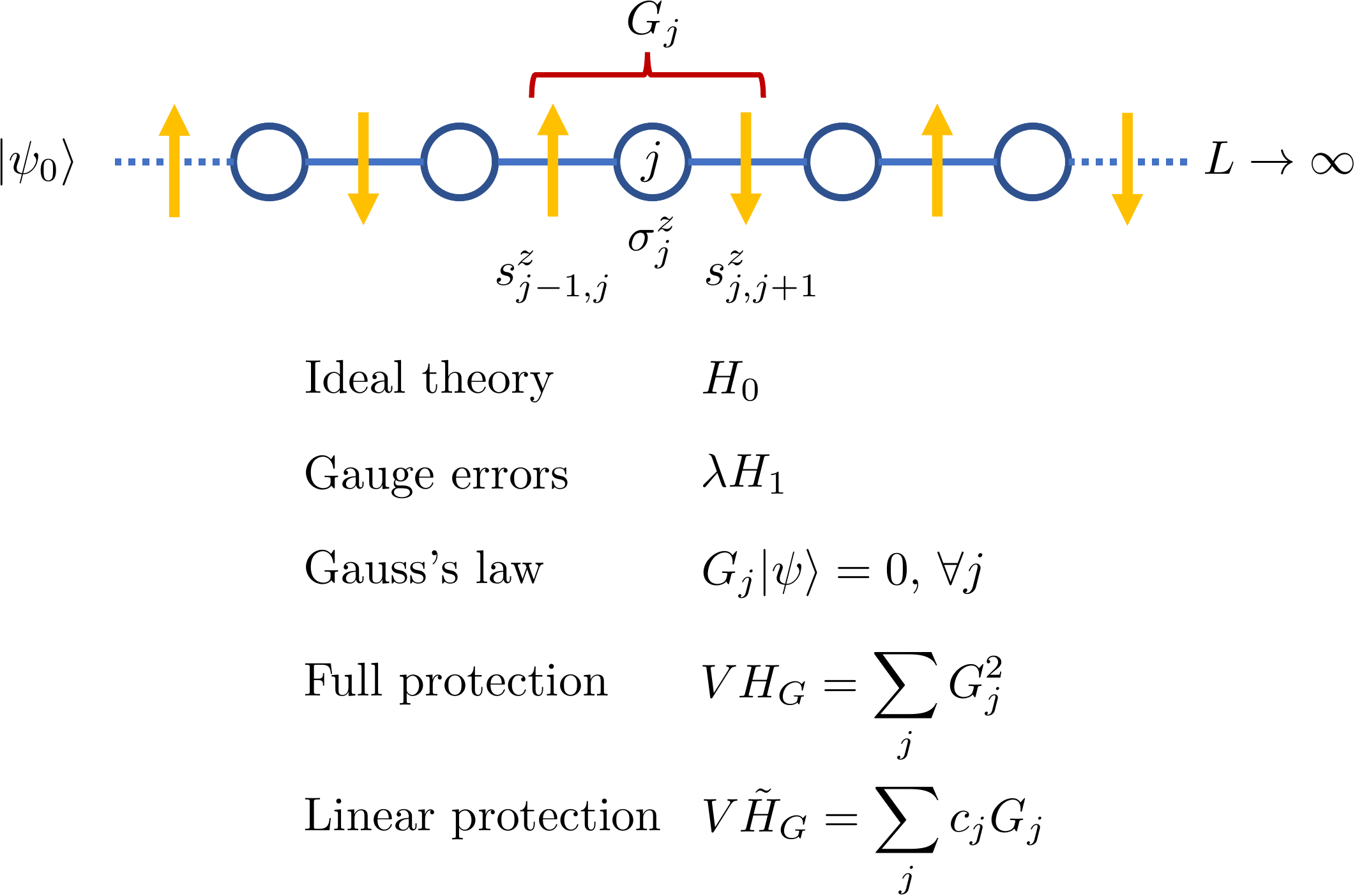}
	\caption{(Color online). Energy protection in lattice gauge theories. The initial gauge-invariant state is quenched by the ``faulty'' gauge theory $H=H_0+\lambda H_1+H_\text{pro}$; the ideal gauge theory $H_0$ is the spin-$S$ $\mathrm{U}(1)$ quantum link model given in Eq.~\eqref{eq:H0}, the gauge-breaking term $\lambda H_1$ describes experimentally relevant local errors given in Eq.~\eqref{eq:error}, and $H_\text{pro}$ is an ``energy penalty'' used to suppress gauge violations and implemented using either \textit{full protection} $H_\text{pro}=VH_G=V\sum_jG_j^2$ or \textit{linear protection} $H_\text{pro}=V\tilde{H}_G=V\sum_jc_jG_j$, with $c_j$ a properly chosen sequence of normalized rational numbers. $G_j$ is the generator of Gauss's law at the local constraint comprised of matter site $j$ and its adjacent links $(j-1,j)$ and $(j,j+1)$, as given in Eq.~\eqref{eq:Gj}. Since $H_0$ is gauge-invariant, it satisfies $[H_0,G_j]=0,\forall j$.	The initial state is chosen in the physical sector $G_j\ket{\psi_0}=0,\forall j$, and such that the matter fields are empty $\sigma_j^z\ket{\psi_0}=-1\ket{\psi_0},\,\forall j$ and with staggered polarization on the links $s_{j,j+1}^z\ket{\psi_0}=(-1)^{j+1}S\ket{\psi_0}$. Here, $\sigma_j^z$ is the Pauli matrix denoting matter occupation at site $j$ and $s_{j,j+1}^z$ is the spin-$S$ matrix depicting the electric field at link $(j,j+1)$. We have checked that the conclusions of our work are independent of the choice of initial state.
	}
	\label{fig:illustration} 
\end{figure}

A convenient framework for the realization of lattice gauge theories is given by quantum link models\cite{Wiese_review,Chandrasekharan1997} (QLMs). In this framework, matter fields (such as electrons and positrons in QED) are represented by fermionic degrees of freedom located at sites of a lattice, while the gauge degree of freedom (such as the electric field) is represented by spins of length $S$ located at the links connecting neighboring lattice sites. One way of protecting gauge symmetry has been proposed in the form of energy penalites.\cite{Zohar2011,Banerjee2012,Zohar2013PRA,Hauke2013,Kuehn2014,Kuno2015,Dutta2017,Kuno2017,Negretti2017,Barros2019,Halimeh2020a,Lamm2020,Halimeh2020e} Reliable gauge invariance in the dynamics of a quantum link spin-$1/2$ lattice gauge theory $H_0$ in the presence of gauge-breaking terms $\lambda H_1$ has been demonstrated numerically in finite systems through the introduction of \textit{full} and \textit{linear} protection terms $VH_G=V\sum_jG_j^2$ and $V\tilde{H}_G=V\sum_jc_jG_j$, respectively. These energetically isolate the target gauge sector in Hilbert space,\cite{Halimeh2020a,Halimeh2020e} where $G_j$ is the generator of Gauss's law at lattice site $j$ and $c_j$ is a sequence of appropriately chosen rational numbers (see Sec.~\ref{sec:lin}); cf.~Fig.~\ref{fig:illustration}. The resulting system has been shown to be analytically connected to the ideal (error-free) theory for a sufficiently large protection- to error-strength ratio $V/\lambda$ for all accessible times in exact diagonalization.\cite{Halimeh2020a} At long times, the gauge violation settles into a steady-state value $\propto\lambda^2/V^2$ in this \textit{controlled-error} regime.

Nevertheless, two interesting questions remain unsettled. How will the gauge protection scheme work in the thermodynamic limit where energetic overlap between different gauge sectors may become unavoidable? And how well does the protection work at larger link spin length $S$? In this work, we show through  infinite matrix product state calculations (iMPS), which work directly in the thermodynamic limit, that at sufficiently large volume-independent protection strength $V$ the gauge violation remains reliably suppressed $\propto\lambda^2/V^2$ up to all accessible evolution times for both full and linear gauge protection. 
Further, we demonstrate analytically that $V$ scales at worst $\sim S^2$ under full protection $VH_G$. Even more, specific realizations can perform decisively better than this rigorous bound. For example, in the scenario considered in this work, the gauge violation is suppressed further with $S$ for fixed values of $\lambda$ and $V$ under both full and linear protection. Moreover, we analytically prove in the case of full protection at sufficiently large volume-independent $V$ that an \textit{adjusted} version of the ideal gauge theory arises and persists until a timescale $\tau_\text{adj}\propto\sqrt{V/V_0^3}$, where the energy scale $V_0$ is roughly given by a linear sum of $\{\lambda,g^2aS^2,\mu,J\}$, which means $V_0\sim S^2$ in the worst scenario. This complements earlier findings\cite{Halimeh2020e} for an adjusted gauge theory in the case of linear protection up to a timescale $\tilde{\tau}_\text{adj}\propto V/(V_0L)^2$.

In the subsequent stages, a renormalized gauge theory dominates lasting up to a timescale $\tau_\text{ren}\propto\exp(V/V_0)/V_0$ for both full and linear protection, with the latter only in the case of a compliant sequence (no such emergent theory exists in the case of a noncompliant sequence; see Sec.~\ref{sec:lin} and Ref.~\onlinecite{Halimeh2020e}). Only beyond this exponentially large timescale, the energy protection can no longer guarantee reliable gauge invariance. It is to be noted, however, that in the case of linear protection with a compliant sequence, $V$ will have to be increased with system size to maintain a given level of gauge fidelity, because the spacing in the compliant sequence $c_j$ decreases with system size (see Sec.~\ref{sec:lin}).

The rest of the paper is organized as follows: We present our main model in Sec.~\ref{sec:model}, where the ideal gauge theory is represented by the one-dimensional spin-$S$ $\mathrm{U}(1)$ quantum link model in the presence of  experimentally relevant gauge-breaking terms. In Sec.~\ref{sec:quench}, we present our numerical and analytic results on quench dynamics of various observables while including either full (Sec.~\ref{sec:full}) or linear (Sec.~\ref{sec:lin}) gauge-protection terms to suppress gauge violations. In Sec.~\ref{sec:timescale}, we summarize the different emergent gauge theories and corresponding timescales. We conclude and discuss future directions in Sec.~\ref{sec:conc}. We complement the main part of the paper with several Appendices. Supporting numerical results are found in Appendix~\ref{sec:support}. The analytics of our paper are found in Appendix~\ref{sec:Abanin} for the Abanin-De Roeck-Ho-Huveneers (ARHH) method,\cite{abanin2017rigorous} Appendix~\ref{sec:constrained} for constrained quantum dynamics,\cite{gong2020universal,gong2020error} and Appendix~\ref{sec:QZE} for the quantum Zeno effect.\cite{facchi2002quantum,facchi2004unification,facchi2009quantum,burgarth2019generalized}

\section{Model}\label{sec:model}
To put our discussion on a formal footing, we consider the paradigmatic one-dimensional spin-$S$ $\mathrm{U}(1)$ quantum link model (QLM) described by the Hamiltonian\cite{Hauke2013,Yang2016,Kasper2017}
\begin{align}\nonumber
H_0=&\,\sum_{j=1}^L\bigg[\frac{-J}{2a\sqrt{S(S+1)}}\big(\sigma^-_js^+_{j,j+1}\sigma^-_{j,j+1}+\text{H.c.}\big)\\\label{eq:H0} 
&+\frac{\mu}{2}\sigma^z_j+\frac{g^2a}{2}\big(s^z_{j,j+1}\big)^2\bigg]\,.
\end{align}
The Pauli matrices $\sigma_j$ represent the matter fields on matter site $j$ with rest mass $\mu$, where $\sigma^z_j$ denotes their occupation and $\sigma^\pm_j$ are the creation and annihilation operators, respectively. $L$ is the total number of matter sites.
The spin-$S$ matrices $s^{x,z}_{j,j+1}$ represent the gauge and electric fields, respectively, with electric-field-ladder operators $s^\pm_{j,j+1}$, on the link $(j,j+1)$. The lattice spacing and gauge coupling are denoted by $a$ and $g$, respectively. The term $\propto J$, which couples the matter and gauge fields, describes the creation or annihilation of an `electron-positron' pair and the concomitant flipping in the intermediate electric field to satisfy Gauss's law, the generator of which is
\begin{align}\label{eq:Gj}
G_j=\frac{(-1)^j}{2}\Big[2\big(s^z_{j-1,j}+s^z_{j,j+1}\big)+\sigma^z_j+1\Big].
\end{align}
We consider the physical (or target) sector as that consisting of the states $\ket{\psi}$ such that $G_j\ket{\psi}=0,\,\forall j$, where Gauss's law is satisfied. The $\mathrm{U}(1)$ QLM is gauge-invariant, i.e., $[H_0,G_j]=0,\,\forall j$. 

In a realistic experiment without infinite fine-tuning, however, there will be unavoidable coherent errors that violate Gauss's law. Even if just perturbative, such errors are crucial to understand and, if possible, control, in order to have a reliable gauge-theory implementation. An experimentally relevant local error term is
\begin{align}\label{eq:error}
\lambda H_1=\lambda\sum_{j=1}^L\bigg[\sigma^-_j\sigma^-_{j+1}+\sigma^+_j\sigma^+_{j+1}+\frac{2s^x_{j,j+1}}{\sqrt{S(S+1)}}\bigg],
\end{align}
of strength $\lambda$, which describes unassisted creation or annihilation of `electron-positron' pairs and unassisted electric-field flipping, either of which will break gauge invariance and ultimately undermine a gauge-theory realization.\cite{Mil2020}

\section{Quench dynamics}\label{sec:quench}
Several proposals have been put forward to protect against gauge-breaking errors such as those in Eq.~\eqref{eq:error},\cite{Zohar2011,Banerjee2012,Zohar2013PRA,Hauke2013,Kuehn2014,Kuno2015,Dutta2017,Kuno2017,Negretti2017,Barros2019,Halimeh2020a,Lamm2020,Halimeh2020e} with one straightforward method that has received a lot of interest being an energy term that penalizes processes away from the physical sector $G_j\ket{\psi}=0,\,\forall j$. 
In the following, we will study quench dynamics in such a scenario. A system initially prepared in a gauge-invariant state $\ket{\psi_0}$ in the target sector ($G_j\ket{\psi_0}=0,\forall j$) is subsequently quenched by a ``faulty'' gauge-theory implementation 
\begin{align}\label{eq:faulty}
H=H_0+\lambda H_1+H_\text{pro},
\end{align}
where $H_\text{pro}$ is the gauge-protection term whose purpose is to suppress gauge violations arising due to $\lambda H_1$. We will employ two variants of this gauge-protection term: the \textit{full protection}\cite{Halimeh2020a} $H_{\rm pro}=VH_G=V\sum_j G_j^2$ (see Sec.~\ref{sec:full}), and the \textit{linear protection}\cite{Halimeh2020e} $H_{\rm pro}=V\tilde{H}_G=V\sum_j c_jG_j$, where $c_j$ is a sequence of rational numbers that we will discuss further in Sec.~\ref{sec:lin}. In the following, we will provide our numerical results calculated in ED and iMPS. For the latter, we find that our most demanding calculations achieve convergence at a bond dimension $\mathcal{D}=200$ and a time-step of $\Delta t=0.005/J$. In our ED implementations, we employ periodic boundary conditions, leading to a system of size $2L$, with $L$ matter sites and $L$ corresponding links. We note, however, that removing the periodic boundary conditions does not qualitatively alter the ED results.\cite{Halimeh2020a}

\subsection{Full protection}\label{sec:full}
A natural way to implement the energy penalty is in the form
\begin{align}\label{eq:FullProtection}
VH_G=V\sum_jG_j^2,
\end{align}
with $V$ the protection strength. We shall refer to this energy-penalty term as \textit{full gauge protection}. 

It has been shown through exact diagonalization (ED) in finite systems\cite{Halimeh2020a} that upon quenching a gauge-invariant initial state in the physical sector with $H=H_0+\lambda H_1+V H_G$, the long-time gauge violation falls in one of two regimes: an uncontrolled-error regime for sufficiently large $\lambda/V$, where the ideal gauge-theory dynamics can no longer be analytically retrieved, or a controlled-error regime for sufficiently small $\lambda/V$ where the gauge violation at long times is suppressed $\propto\lambda^2/V^2$, and the gauge-theory dynamics perturbatively connects to that of the ideal case. The aim of this section is to analyze the persistence of this protective power through numerical calculations in the thermodynamic limit and through additional analytic bounds.

\begin{figure*}[htp]
	\centering
	\includegraphics[width=.48\textwidth]{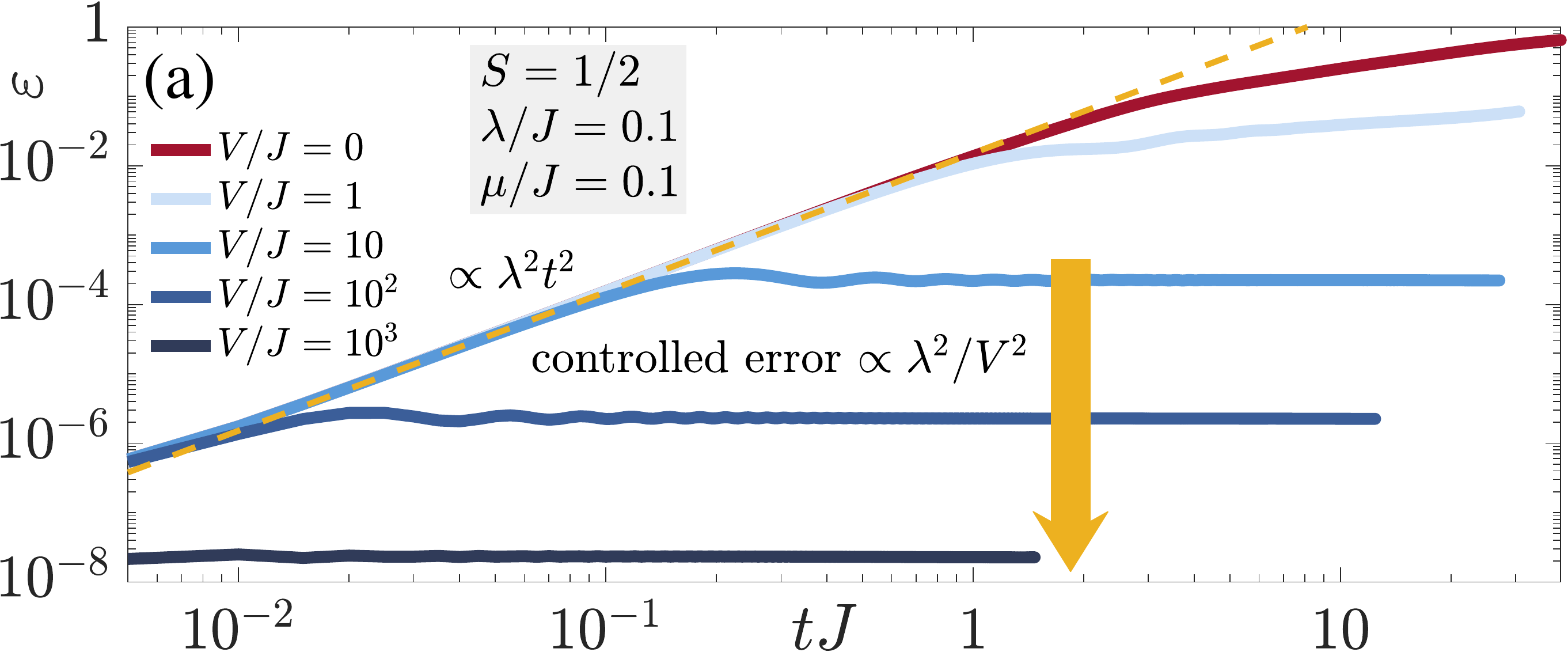}\quad
	\includegraphics[width=.48\textwidth]{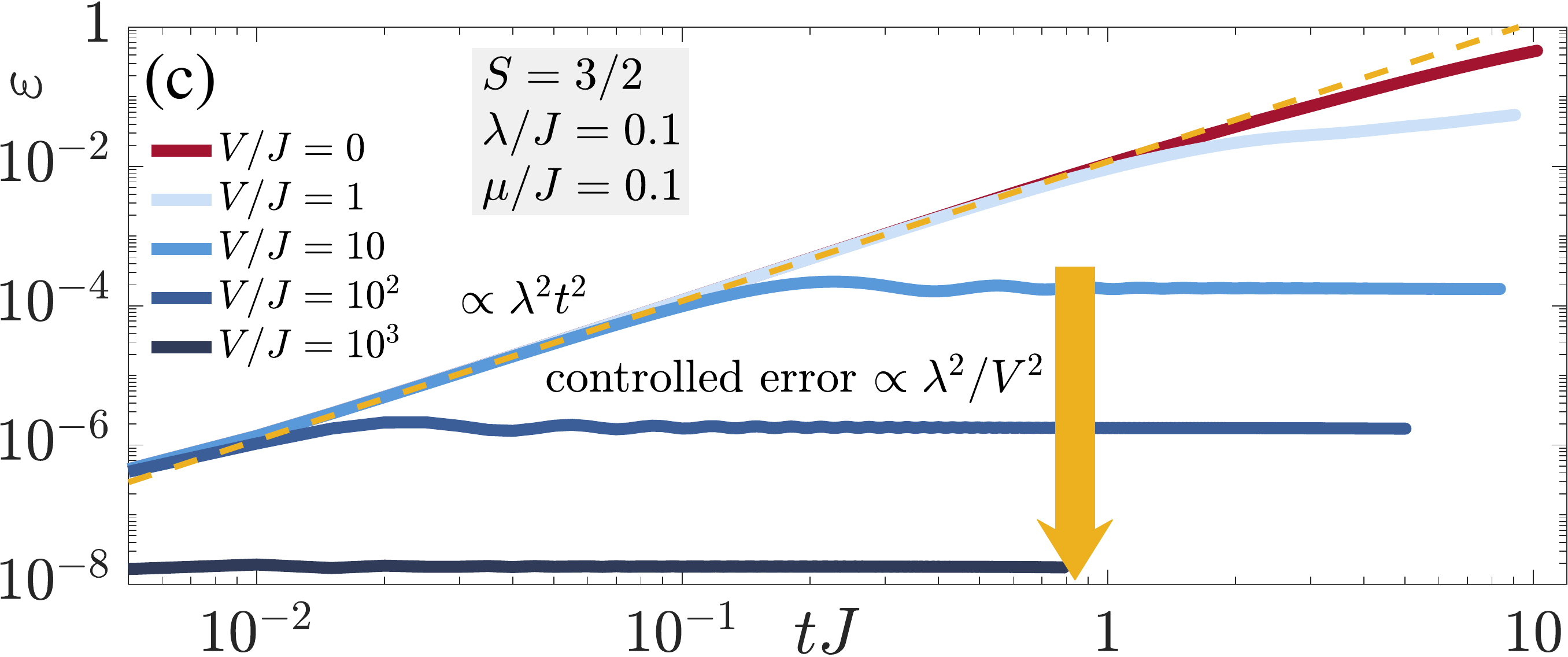} \\
	\includegraphics[width=.48\textwidth]{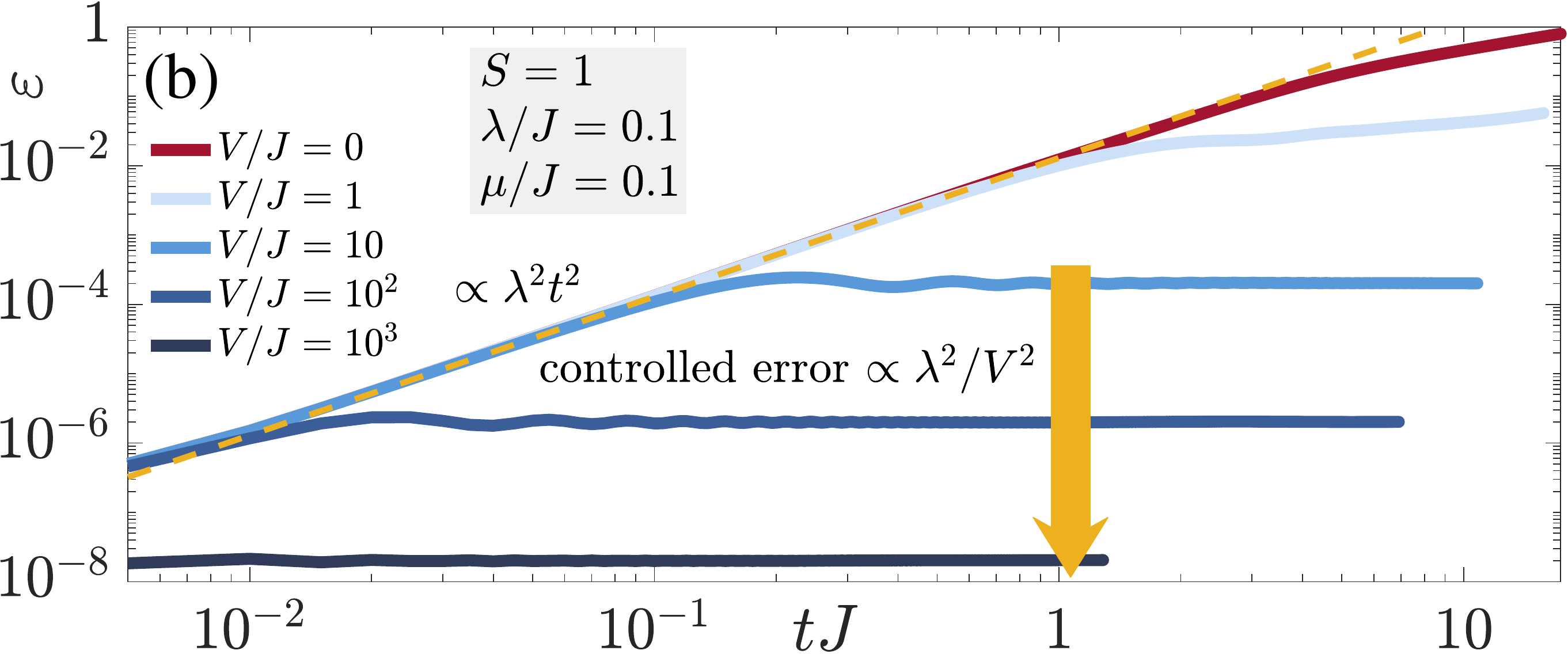}\quad
	\includegraphics[width=.48\textwidth]{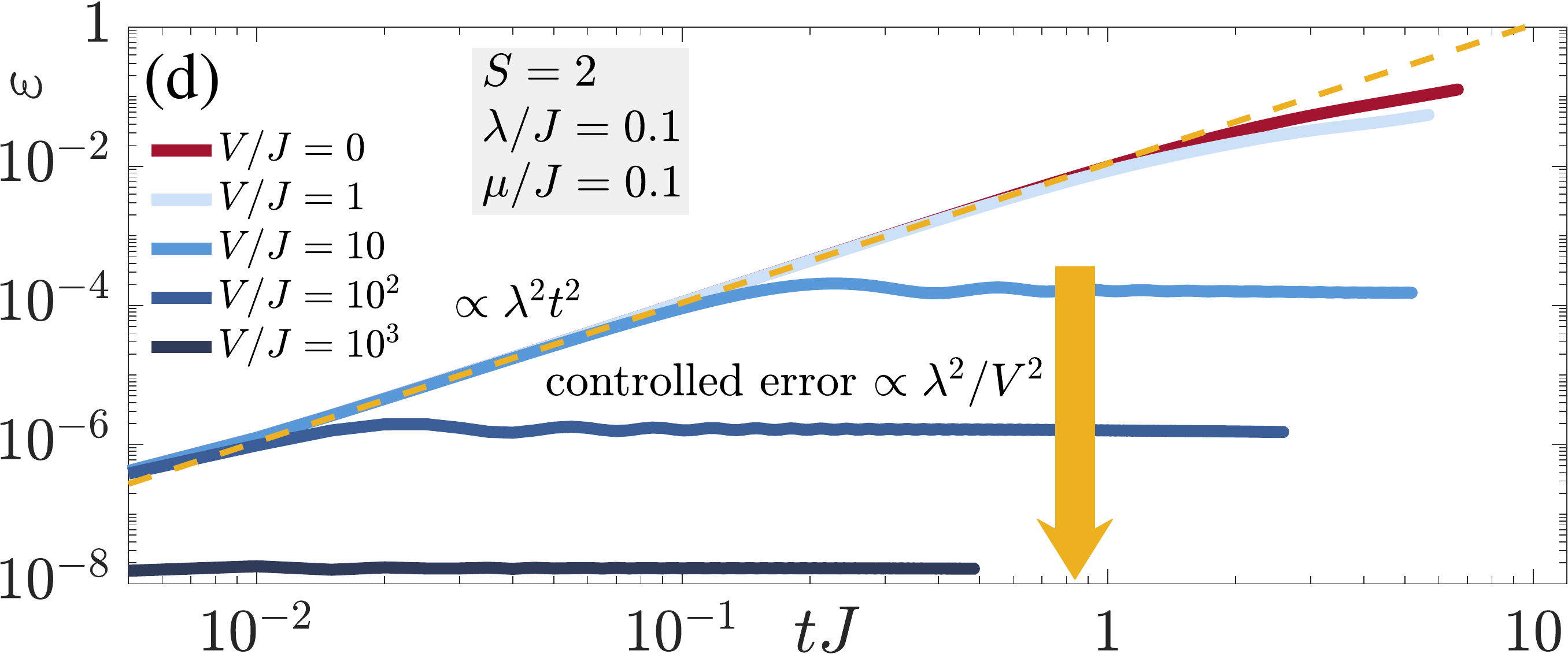}
	\caption{(Color online). 
		Gauge violation as given in Eq.~\eqref{eq:violation} as function of gauge-spin length $S$ [(a) $S=1/2$, (b) $S=1$, (c) $S=3/2$, and (d) $S=2$], computed in the thermodynamic limit using iMPS. 
		The initial state, outlined in Fig.~\ref{fig:illustration}, is quenched with Hamiltonian $H_0+\lambda H_1+V H_G$, i.e., using the ``full'' gauge protection given in Eq.~\eqref{eq:FullProtection}. Errors, generated by the gauge-breaking term $\lambda H_1$ of Eq.~\eqref{eq:error}, which may be present in typical ultracold-atom implementations of the $\mathrm{U}(1)$ quantum link model $H_0$ given in Eq.~\eqref{eq:H0}, remain reliably suppressed $\propto\lambda^2/V^2$ for the evolution times accessible in iMPS, regardless of $S$. The plateau $\propto\lambda^2/V^2$ occurs at the timescale $\propto1/V$ at which the protection term begins to dominate.
		Here, we have set $a=1/2$, $g=\sqrt{J}$, $\lambda=0.1J$, and $\mu=0.1J$, and scanned over the protection-strength values $V/J=0,\,1,\,10,\,10^2,\,10^3$, although our conclusions are valid also for other values.
	}
	\label{fig:FullProtection} 
\end{figure*}

Using the iMPS technique based on the time-dependent variational principle (TDVP),\cite{Haegeman2011,Haegeman2016,Vanderstraeten2019} we calculate the quench dynamics under the Hamiltonian $H$ of Eq.~\eqref{eq:faulty} starting in the initial state $\ket{\psi_0}$ shown in Fig.~\ref{fig:illustration}. For our results, we have chosen $a=1/2$, $g=\sqrt{J}$, $\lambda=0.1J$, and $\mu=0.1J$, although we have checked that our conclusions remain valid for different initial states and parameter values. 

The ensuing dynamics of the temporally averaged gauge violation
\begin{align}\label{eq:violation}
\varepsilon(t)=\frac{1}{Lt}\int_0^t ds\,\sum_{j=1}^L\bra{\psi_0}e^{iHs}G_j^2e^{-iHs}\ket{\psi_0},
\end{align}
is shown in Fig.~\ref{fig:FullProtection} for various lengths $S$ of the link spin. The violation grows $\propto\lambda^2t^2$ at short times, in agreement with time-dependent perturbation theory (TDPT).\cite{Halimeh2020a} Whereas in the unprotected ($V=0$) case the gauge violation grows rapidly, at sufficiently large $V$ it is suppressed $\propto\lambda^2/V^2$ at long times, settling into a plateau beginning at a timescale $t\propto1/V$ at sufficiently large $V$. This is the timescale at which the protection term begins to dominate. The gauge violation then remains at this plateau throughout all the accessible evolution times in iMPS. We see that for fixed $\lambda$ and $V$, the gauge-violation plateau actually slightly decreases with increasing $S$, although this may be specific to our error term in Eq.~\eqref{eq:error}. Indeed, the protection strength $V$ can be shown analytically to scale $\sim S^2$ in the worst case; cf.~Appendices~\ref{sec:Abanin} and~\ref{sec:constrained}. Nevertheless, the apparent small dependence of $V$ on $S$ in the case of experimentally relevant local gauge-breaking errors such as those of Eq.~\eqref{eq:error} is very encouraging for ongoing experiments seeking to approach the Kogut-Susskind limit (lattice QED) by achieving larger link spin lengths in gauge-theory implementations.\cite{Mil2020}

\begin{figure}[htp]
	\centering
	\includegraphics[width=.48\textwidth]{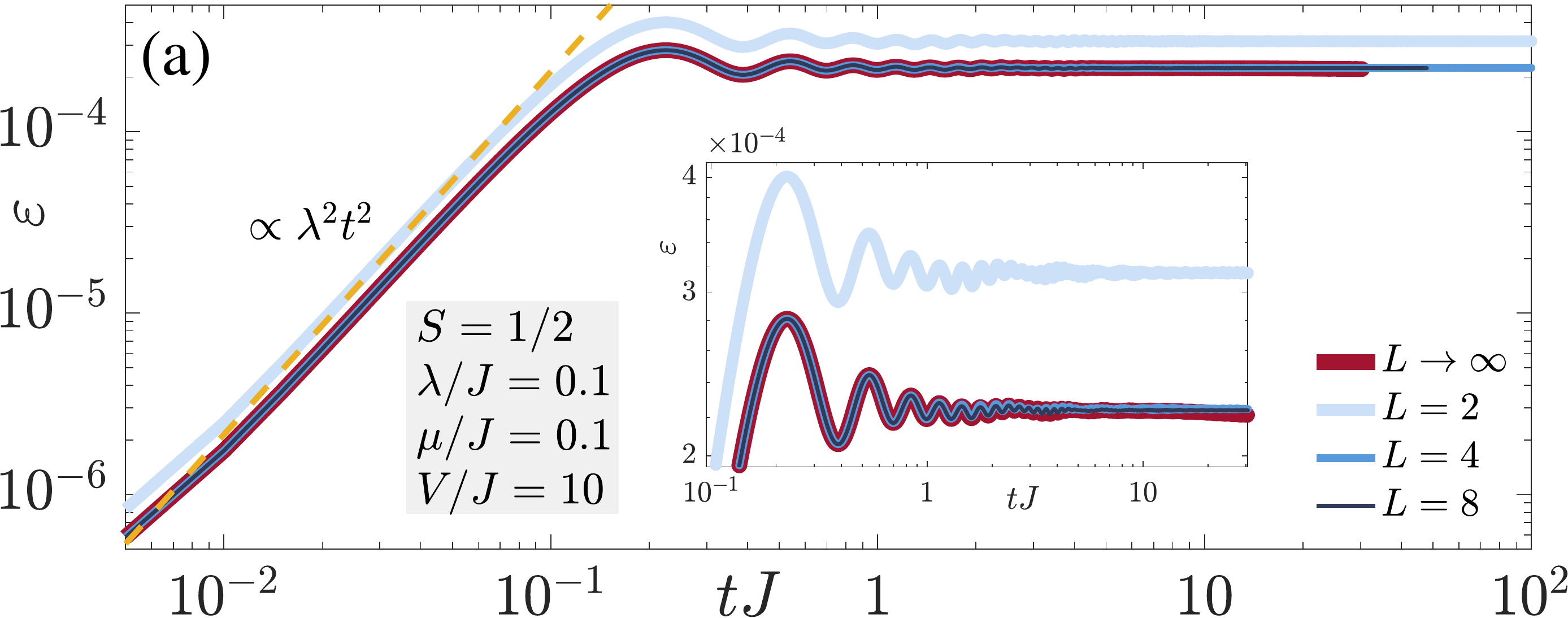}\\
	\vspace{1.5mm}
	\includegraphics[width=.48\textwidth]{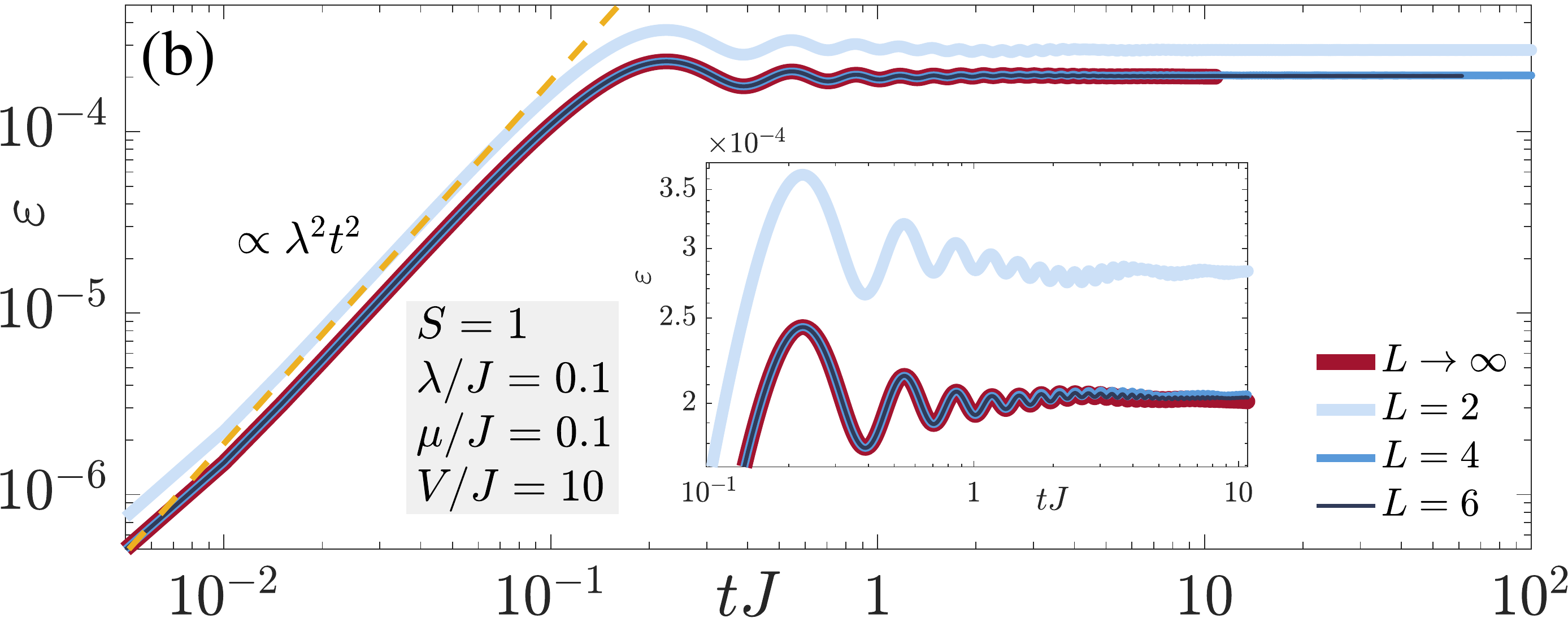}
	\caption{(Color online). Finite-size behavior of the gauge violation at $\lambda=0.1J$ and $V=10J$, for the link spin lengths (a) $S=1/2$ and (b) $S=1$. The gauge violation for finite-size chains is larger than its counterpart in the thermodynamic limit (see legends). As can be seen in the insets, convergence relative to system size is very fast, with the gauge violation for $L=4$ matter sites already approaching the results in the thermodynamic limit $L\to\infty$. We have checked that this behavior persists for other values of $\lambda$ and $V$. The dynamics in the thermodynamic limit is calculated in iMPS, while for finite systems in ED.}
	\label{fig:FSSfull} 
\end{figure}

To analyze the finite-size behavior of the gauge violation, we show in Fig.~\ref{fig:FSSfull}(a,b) for $S=1/2$ and $1$, respectively, at $\lambda=0.1J$ and $V=10J$, the gauge-violation dynamics at several finite values of $L$, calculated in ED, along with the thermodynamic limit $L\to\infty$, calculated in iMPS. Convergence to the latter is very fast, with deviations from the iMPS results becoming insignificant already with only $L=4$ matter sites (see insets). This is remarkable especially from an experimental perspective since state-of-the-art quantum-simulator implementations of lattice gauge theories are of the order of a few dozen matter sites,\cite{Yang2020} and our results indicate that this already captures the thermodynamic limit in the behavior of the gauge violation.

Our iMPS results show that the gauge violation in the thermodynamic limit will still quickly settle into a plateau at sufficiently large $V/\lambda$, but they do not tell us much about whether such a plateau will last indefinitely given the inaccessibility of longer evolution times in iMPS. However, it has been analytically proven that in the case of a controlled suppression of the gauge violation in the presence of local gauge-breaking terms as those in Eq.~\eqref{eq:error}, full protection at a volume-independent strength $V$ will give rise to a renormalized gauge theory, where the gauge violation is upper bounded $\propto\lambda/V$ and which lasts up to a timescale $\tau_\text{ren}\propto\exp(V/V_0)/V_0$;\cite{abanin2017rigorous,Halimeh2020e}cf.~Appendix~\ref{sec:Abanin}. As such, we cannot rule out that the gauge violation will leave the plateau $\propto\lambda^2/V^2$ appearing in Fig.~\ref{fig:FullProtection} and rise to some larger value with upper bound $\propto\lambda/V$ at longer evolution times $t<\tau_\text{ren}$ that are not accessible to iMPS. Nevertheless, as the iMPS data in Figs.~\ref{fig:FullProtection} and~\ref{fig:FSSfull} show, for times that are accessible to current quantum-simulation experiments\cite{Goerg2019,Schweizer2019,Mil2020,Yang2020} the gauge violation in the thermodynamic limit remains suppressed $\propto\lambda^2/V^2$, same as in the case of finite systems.\cite{Halimeh2020a} Indeed, for example for the experimentally feasible\cite{Halimeh2020a} parameter values $\lambda=0.1J$ and $V=10J$, which lie in the controlled-error regime, the largest evolution times achieved in iMPS are $t_\text{max}\approx10/J$. Setting $J\sim20-30\,\mathrm{Hz}$, which is within the typical range of values used in current ultracold-atom realizations of lattice gauge theories,\cite{Yang2020} leads to $t_\text{max}\sim50-80\,\mathrm{ms}$. This is a typical lifetime of modern large-size lattice-gauge-theory implementations in ultracold-atom setups,\cite{Yang2020} which means that our conclusions from iMPS can be readily checked in such experiments.

\begin{figure*}[htp]
	\centering
	\includegraphics[width=.48\textwidth]{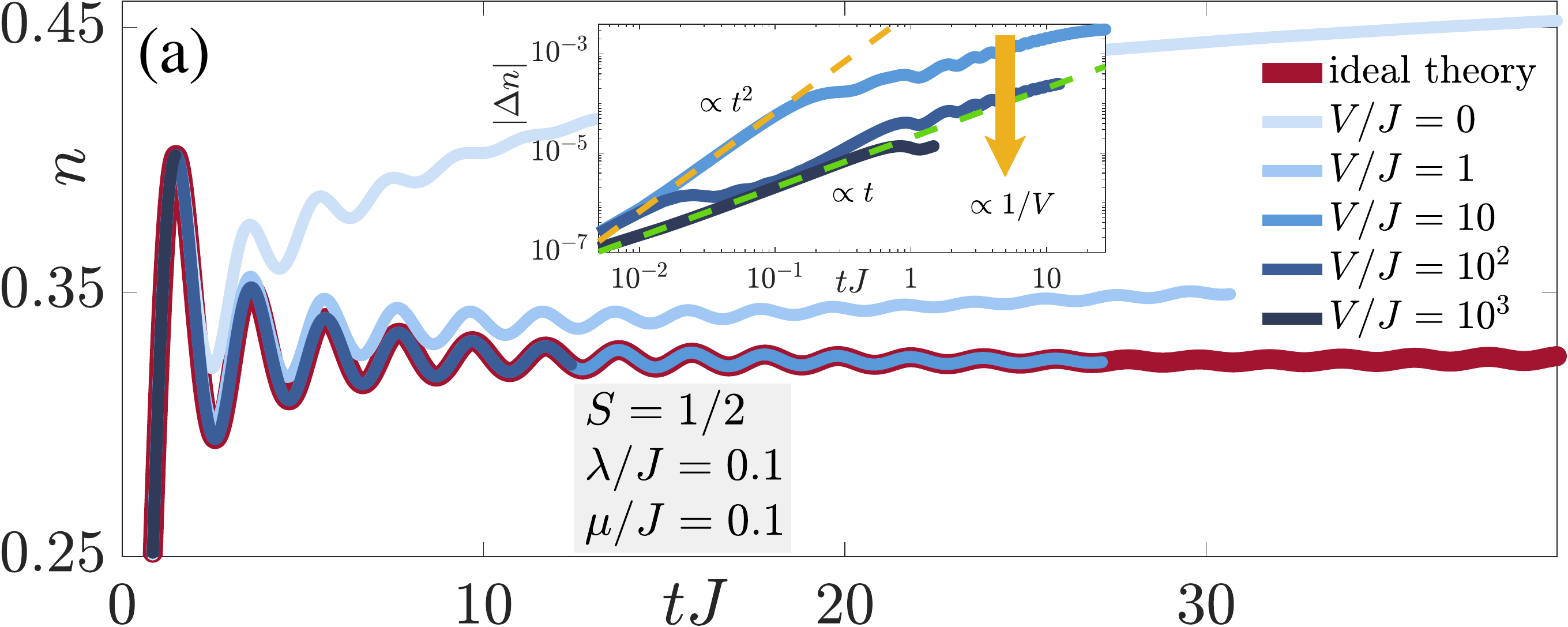}\quad
	\includegraphics[width=.48\textwidth]{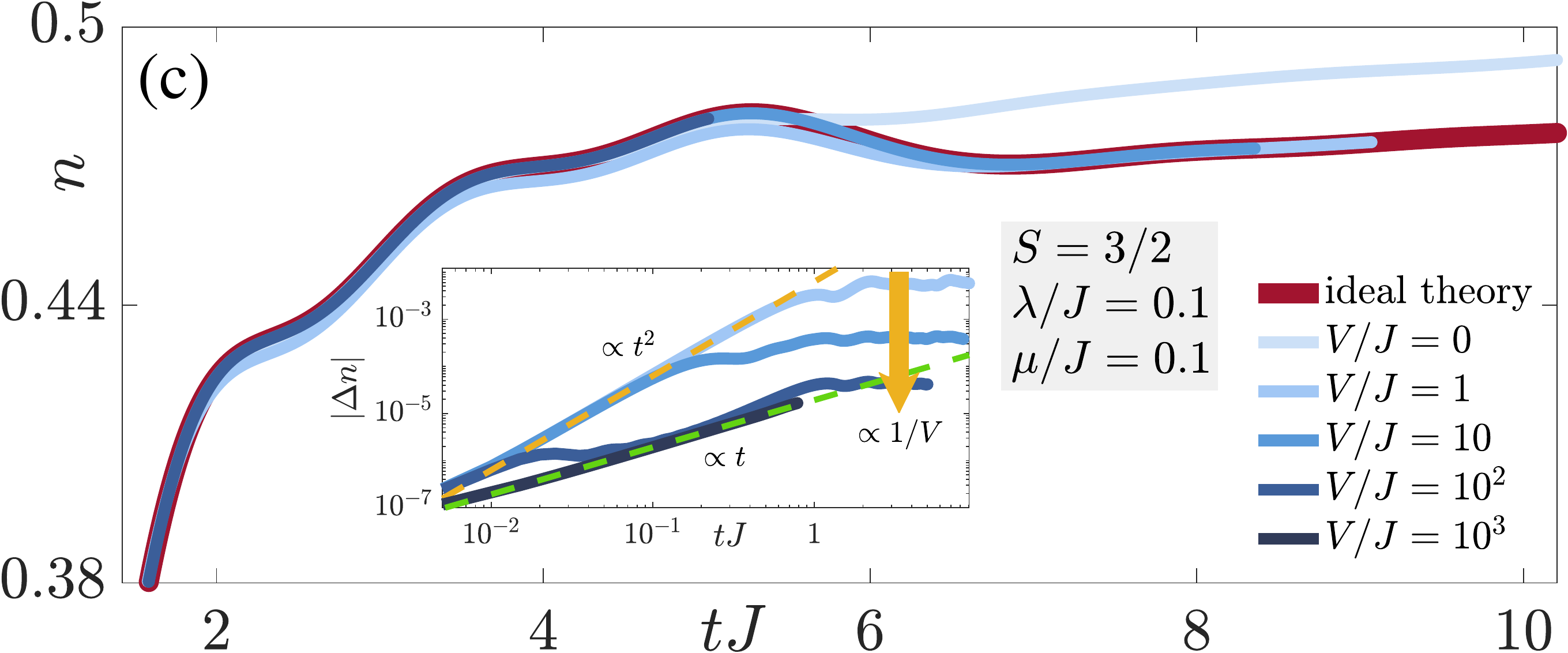} \\
	\vspace{1.5mm}
	\includegraphics[width=.48\textwidth]{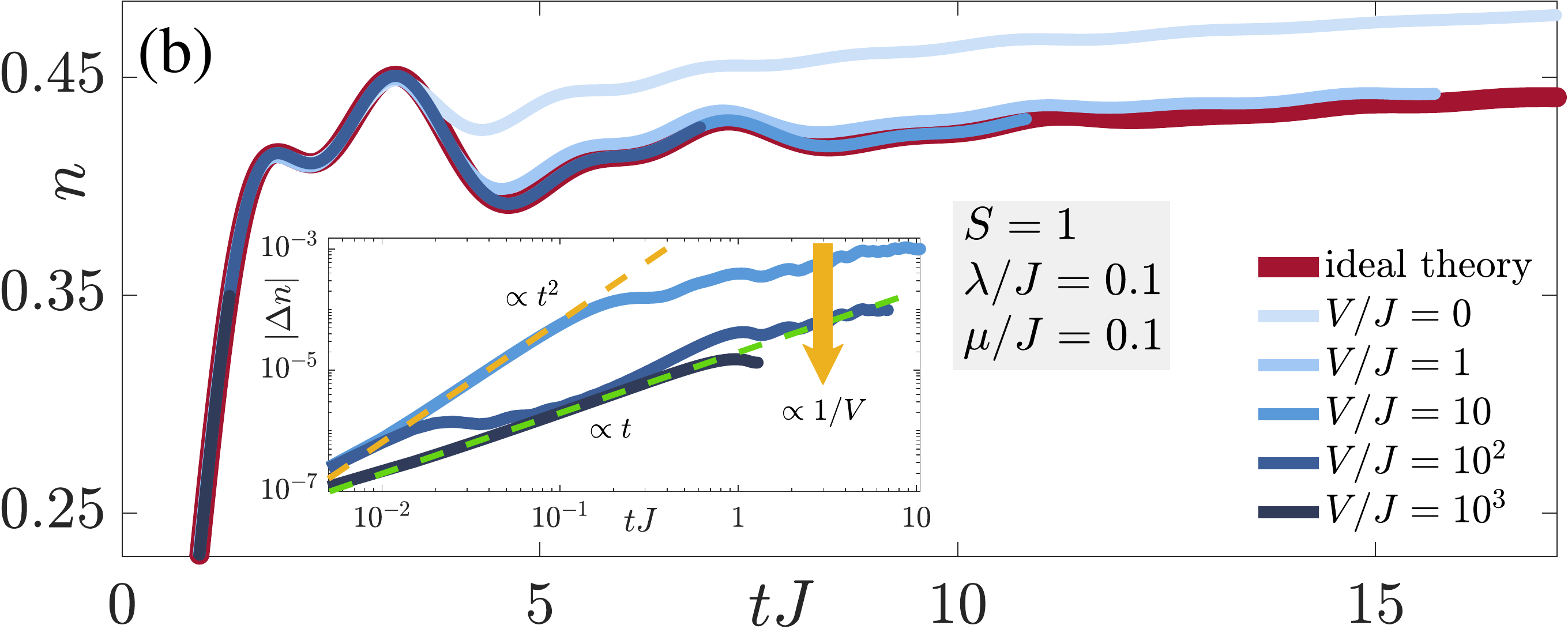}\quad
	\includegraphics[width=.48\textwidth]{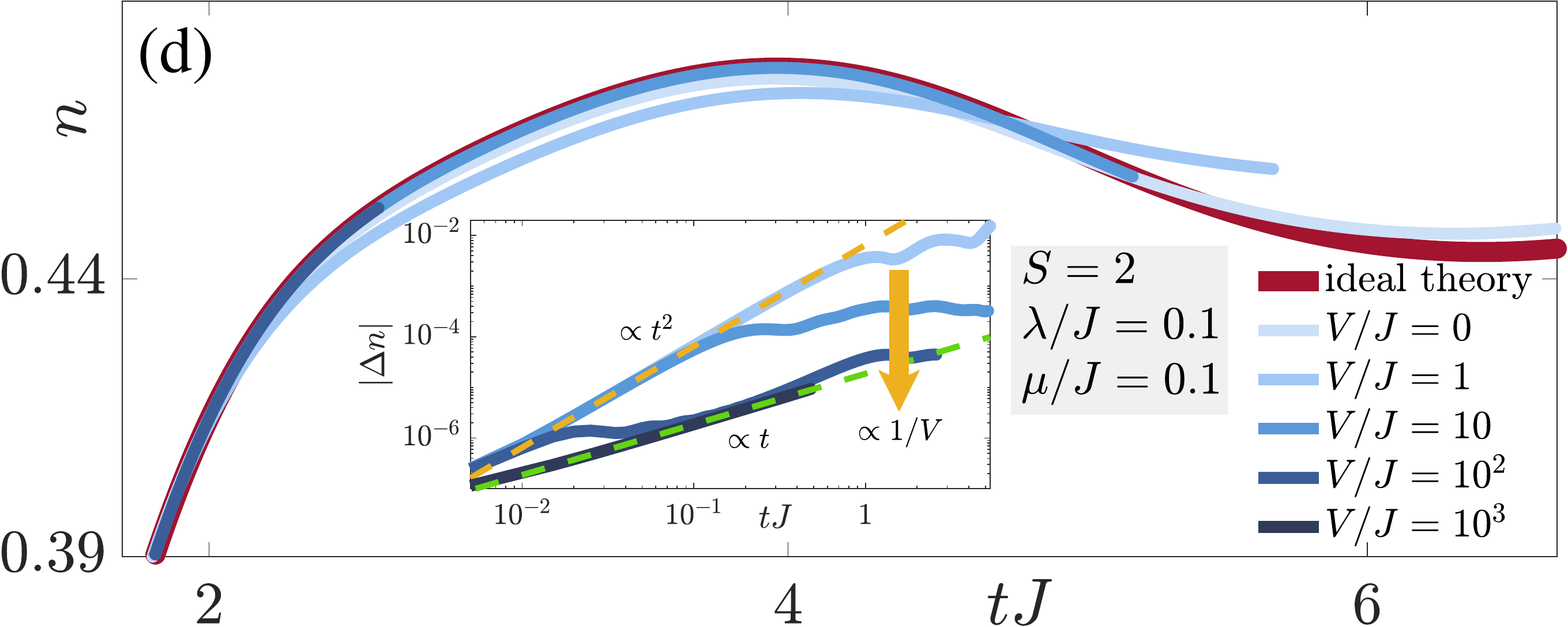}
	\caption{(Color online). 
		Deviation of particle density from ideal dynamics due to gauge-breaking terms.
		Plotted is the temporally averaged particle density for gauge-breaking strength $\lambda/J=0$ (ideal theory; red line), and $\lambda/J=0.1$ with full protection at various strengths $V$ (lines in shades of blue) for link spin lengths (a) $S=1/2$, (b) $S=1$, (c) $S=3/2$, and (d) $S=2$.
		The deviation from the ideal theory is rapidly suppressed as $V$ increases, at least up to the evolution times accessible by iMPS. With increasing $V$, the short-time scaling diminishes from $\propto t^2$ (the theoretical upper bound) to a milder scaling $\propto t$, at a timescale $\propto1/V$ when the protection term dominates the dynamics. 
	}
	\label{fig:FullProtection_sigmaz} 
\end{figure*}

The timescales over which the gauge symmetry can be protected are worthy of a more detailed discussion. The ideal gauge theory $H_0$ is perturbed by the gauge-breaking term $\lambda H_1$, against which we protect with a strong $VH_G$. 
This scenario enables us to demonstrate the existence of an emergent gauge theory that is perturbatively connected to $H_0$. In particular, the frameworks of constrained quantum dynamics\cite{gong2020error,gong2020universal} and the Abanin-De Roeck-Ho-Huveneers (ARHH) method\cite{abanin2017rigorous} prove to yield useful insight. 

First, let us define the adjusted gauge theory $H_\mathrm{adj}=H_0+\lambda\mathcal{P}_0H_1\mathcal{P}_0$, which preserves gauge invariance exactly and includes those terms of $\lambda H_1$ that act purely within the physical sector, denoted by the projector $\mathcal{P}_0$. Now, after quenching a gauge-invariant initial state in the target sector with $H=H_0 + \lambda H_1 + V H_G$, the large scale $V$ restricts the dynamics of local observables to the one generated by $H_\mathrm{adj}$. Within the framework of constrained quantum dynamics, \cite{gong2020error,gong2020universal} one can show the deviation for the local observable $O$, to be bounded from above as  
\begin{align}
	\label{eq:constraineddeviation}
	\big\lvert\langle e^{iHt}Oe^{-iHt} - e^{iH_\text{adj}t}Oe^{-iH_\text{adj}t}\rangle\big\rvert \leq \Delta_\text{adj},
\end{align}
with $\Delta_\text{adj}\sim t^2V_0^3/V$. This polynomial error bound is thus ensured up to a fractional timescale $\tau_\text{adj}\propto\sqrt{V/V_0^3}$. An important feature of this bound is its volume independence, providing an analytic proof complementary to Ref.~\onlinecite{Halimeh2020e} that energy protection of gauge symmetry can work in the thermodynamic limit. The exact form of $\Delta_\text{adj}$ and other derivational details are included in Appendix~\ref{sec:constrained}.

This error bound also restricts the deviation of Gauss's law from the ideal value of $0$. As can be seen in Fig.~\ref{fig:FullProtection}, concrete physical scenarios can remain considerably below this upper bound. In fact, we observe a timescale $\propto1/V<\tau_\text{adj}$ at which the gauge violation plateaus at a constant value $\propto\lambda^2/V^2$, the value that can also be derived in degenerate perturbation theory.\cite{Halimeh2020a} The timescale $\propto1/V$ is when the protection term $H_\text{pro}$ starts to dominate, and therefore the initial gauge violation growth $\propto\lambda^2 t^2$ due to the error term as calculated in TDPT\cite{Halimeh2020a} is suppressed. The gauge violation then remains at that value to times larger than $\tau_\text{adj}$. 
This favorable behavior can be explained within the ARHH framework.\cite{abanin2017rigorous} Namely, one can show the existence of a renormalized gauge theory \cite{Halimeh2020e} that exists up to a timescale $\tau_\text{ren}\propto\exp(V/V_0)/V_0$. This framework bounds the gauge violation within $\lambda/V$, as derived in Ref.~\onlinecite{Halimeh2020e} and also summarized in Appendix~\ref{sec:Abanin}.
\begin{figure}[htp]
	\centering
	\includegraphics[width=.48\textwidth]{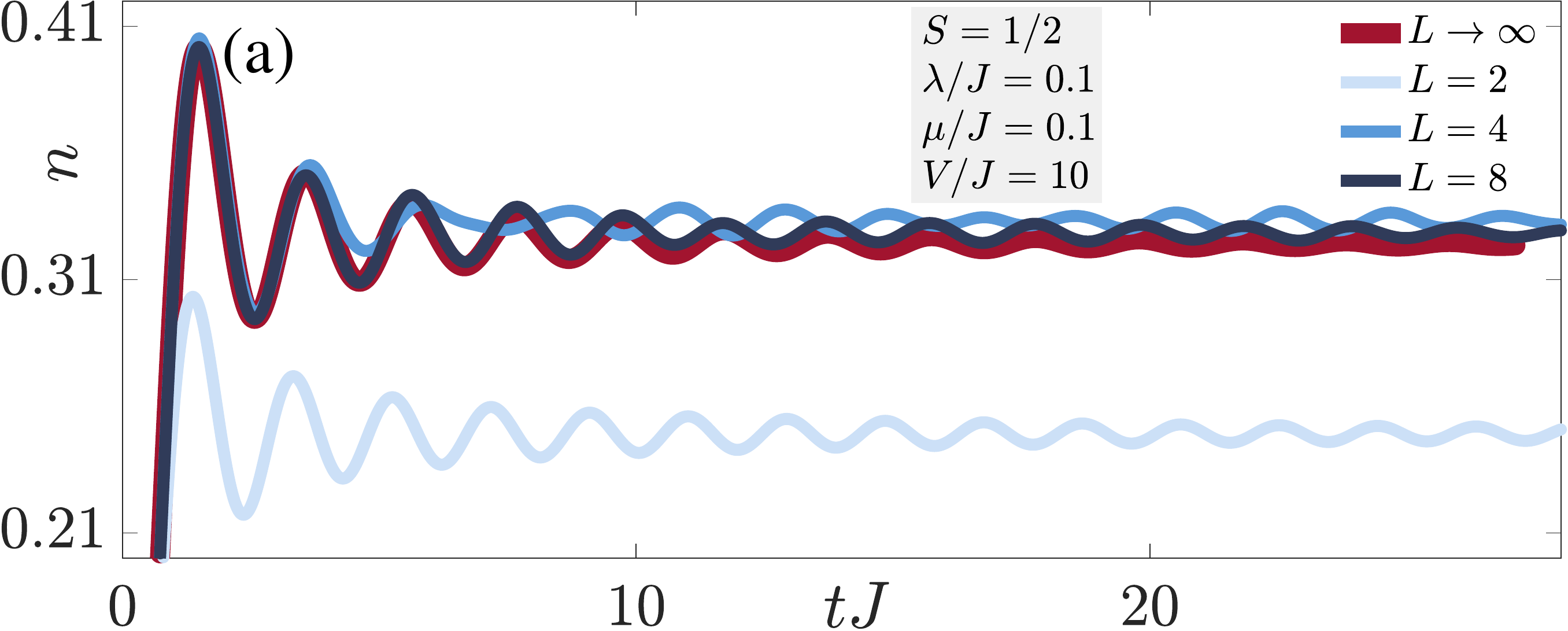}\\
	\vspace{1.5mm}
	\includegraphics[width=.48\textwidth]{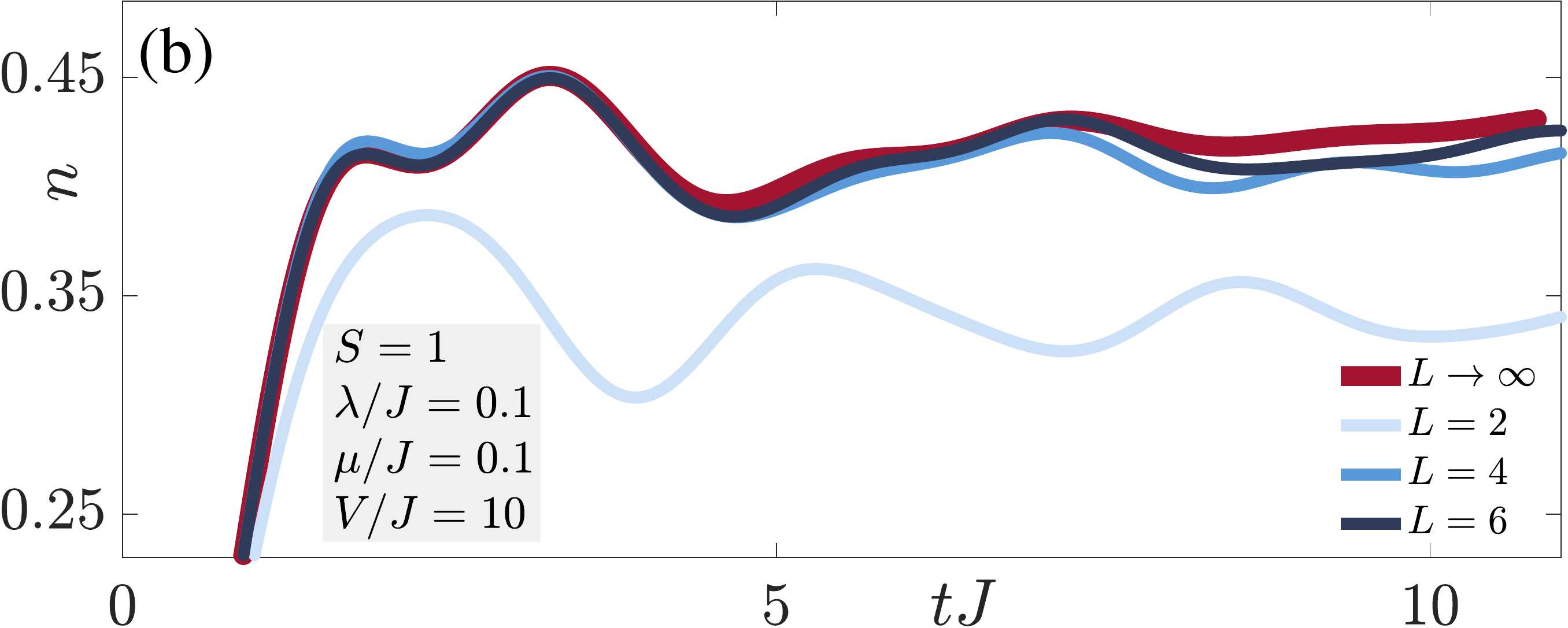}
	\caption{(Color online). Finite-size behavior of the particle density under full gauge protection. Parameters are $\lambda=0.1J$, $V=10J$, and quantum link spin length (a) $S=1/2$ and (b) $S=1$. As in the case of the gauge violation in Fig.~\ref{fig:FSSfull}, the result quickly approaches its value in the thermodynamic limit already at small system sizes. Our results hold for other values of the microscopic parameters. The dynamics in the thermodynamic limit is calculated in iMPS, while for finite systems in ED.
	}
	\label{fig:FSSfull_sigmaz} 
\end{figure}

As these discussions illustrate, there exist powerful frameworks to derive analytic bounds for the gauge violation, which realistic situations can even significantly undershoot.

It is instructive to look at other local observables where the timescale $\propto1/V$ does not prominently appear as it does in the gauge violation, and where we can therefore more clearly see the features of the adjusted gauge theory $H_\text{adj}=H_0+\lambda\mathcal{P}_0H_1\mathcal{P}_0$. Indeed, even though at times longer than the timescale $\propto1/V$ the protection term dominates, its effect in general local observables may not be as apparent as in the plateauing behavior in the gauge violation of Fig.~\ref{fig:FullProtection}. This in turn would allow us to better estimate how well the dynamics of the faulty theory $H$ reproduces the one of the adjusted gauge theory. The particular gauge-breaking error described by Eq.~\eqref{eq:error} is characterized by $\mathcal{P}_0H_1\mathcal{P}_0=0$, and so the adjusted gauge theory is $H_0$ itself. As such, we expect local observables after the quench with the faulty theory $H$ to follow the ideal-theory dynamics for sufficiently large $V$, up to an error that according to Eq.~\eqref{eq:constraineddeviation} is at most $\propto t^2V_0^3/V$. To check this expectation, we calculate in iMPS the dynamics of the temporally averaged particle density
\begin{align}\label{eq:density}
	n&=\frac{1}{2}+\frac{1}{2Lt}\int_0^t ds\,\sum_{j=1}^L\bra{\psi_0}e^{iHs}\sigma_j^ze^{-iHs}\ket{\psi_0},
\end{align}
which we present in the main text, in addition to the temporally averaged absolute value of the normalized (by spin length $S$) electric field
\begin{align}\label{eq:electric}
E&=\frac{1}{SLt}\int_0^t ds\,\Big\lvert\sum_{j=1}^L(-1)^j\bra{\psi_0}e^{iHs}s_j^ze^{-iHs}\ket{\psi_0}\Big\rvert,
\end{align}
which we relegate to Appendix~\ref{sec:electric}. The same conclusions are drawn from both these local observables.

Figure~\ref{fig:FullProtection_sigmaz} shows the time evolution of the temporally averaged particle density at gauge-breaking strength $\lambda=0.1J$ and various protection strengths $V$ for several values of the link spin length $S$, and compares the result to the corresponding ideal gauge-theory dynamics propagated by only $H_0$. Once in the controlled-error regime ($V\gtrsim10J$), the particle density agrees well with that of the ideal theory ($\lambda=V=0$; thick red line) up to an error upper bound $\propto t^2V_0^3/V$ at short times, in support of our analytic predictions. Indeed, as seen in the insets, the deviation $\lvert\Delta n\rvert$ between the particle density due to a quench by Eq.~\eqref{eq:faulty} and that due to the ideal theory $H_0$ is suppressed at later times as $1/V$ while growing slower than $t^2$, i.e., substantially more benign than our analytically derived error upper bound. In priniple, this comparison can also be carried out with the renormalized gauge theory as a reference for the error upper bound. However, the form of the Hamiltonian for the renormalized gauge theory is in general not known, rendering such a comparison unfeasible. 

The finite-size behavior of the particle density is shown in Fig.~\ref{fig:FSSfull_sigmaz} for two values of the link spin length $S=1/2$ and $1$, and the quench parameters $\lambda=0.1J$ and $V=10J$, although we have checked that our conclusions hold for other values of the microscopic parameters. Similar to the case of the gauge violation in Fig.~\ref{fig:FSSfull}, the finite-size behavior of the particle density calculated in ED approaches already at a few matter sites the dynamics in the thermodynamic limit within the evolution times accessible in iMPS. This is particularly encouraging news from an experimental perspective, because it shows that modern ultracold-atom implementations of lattice gauge theories, which now reach a few dozens of matter sites,\cite{Yang2020} should be able to faithfully achieve the thermodynamic behavior of gauge-theory dynamics at least up to timescales $\propto\sqrt{V/V_0^3}$.

Finally, it is worth mentioning that the timescale $\propto1/V$, though not evident in the dynamics of the particle density itself, does actually appear in the deviation of the particle density from its dynamics under the ideal theory, as can be seen in the insets of Fig.~\ref{fig:FullProtection_sigmaz}.

\begin{figure*}[htp]
	\centering
	\includegraphics[width=.48\textwidth]{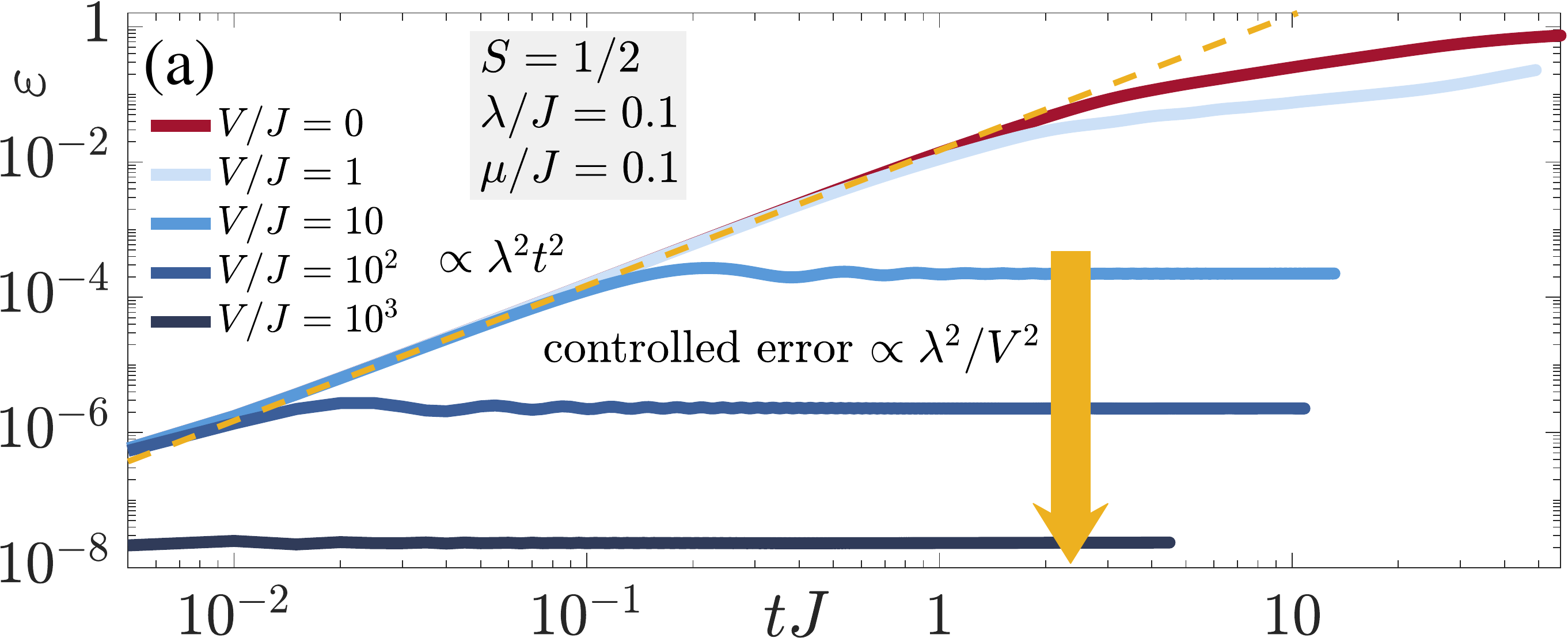}\quad
	\includegraphics[width=.48\textwidth]{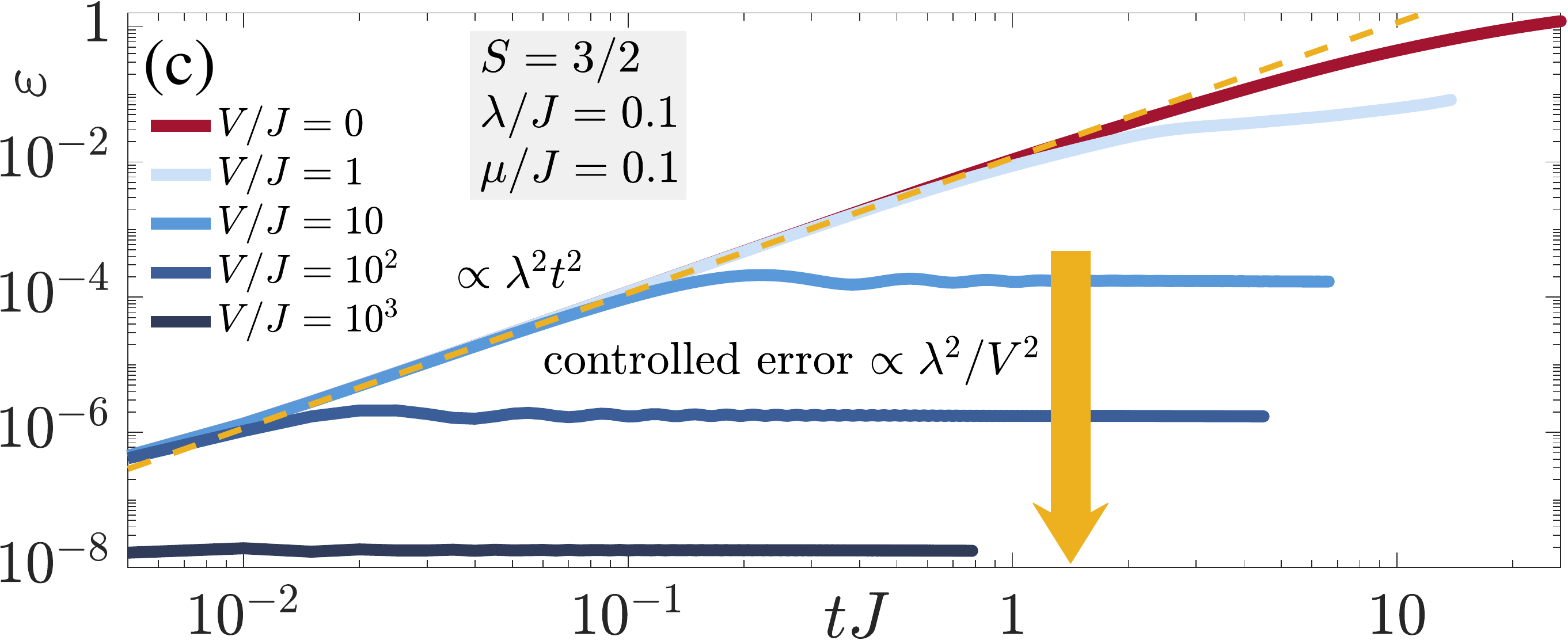} \\
	\vspace{1.1mm}
	\includegraphics[width=.48\textwidth]{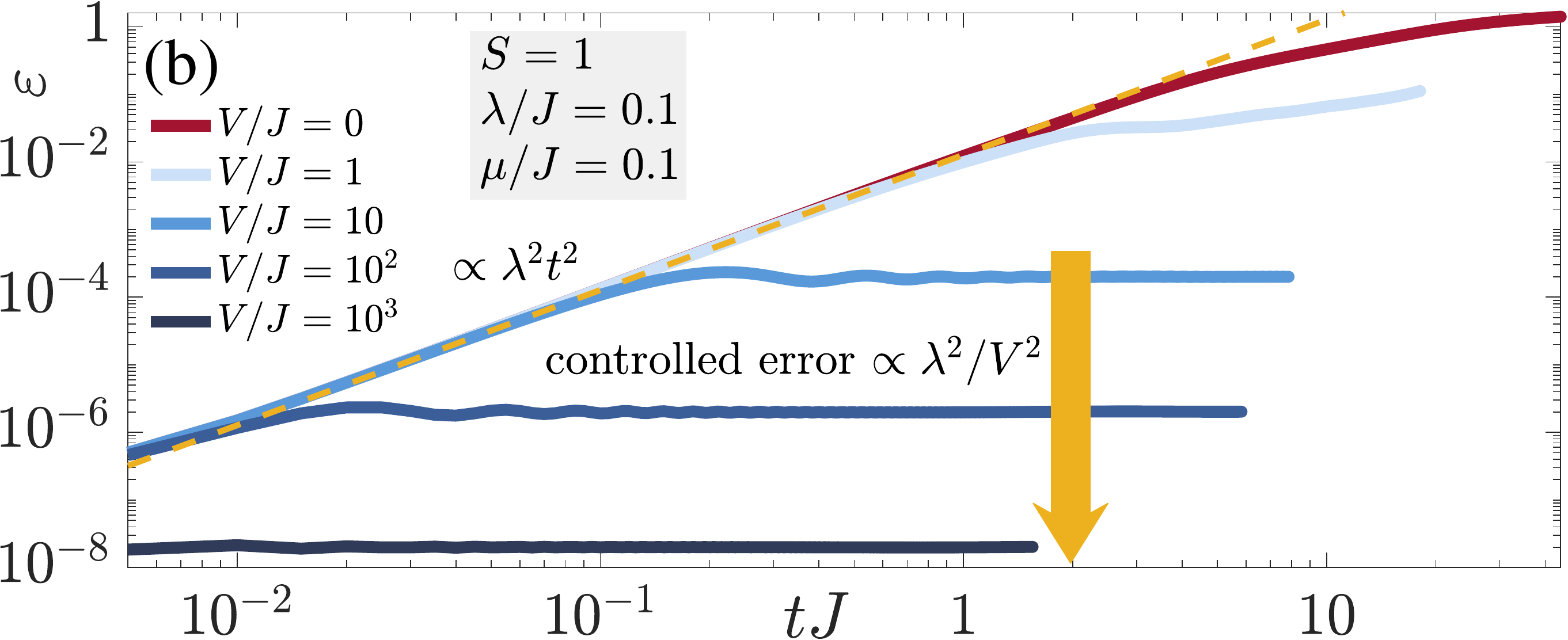}\quad
	\includegraphics[width=.48\textwidth]{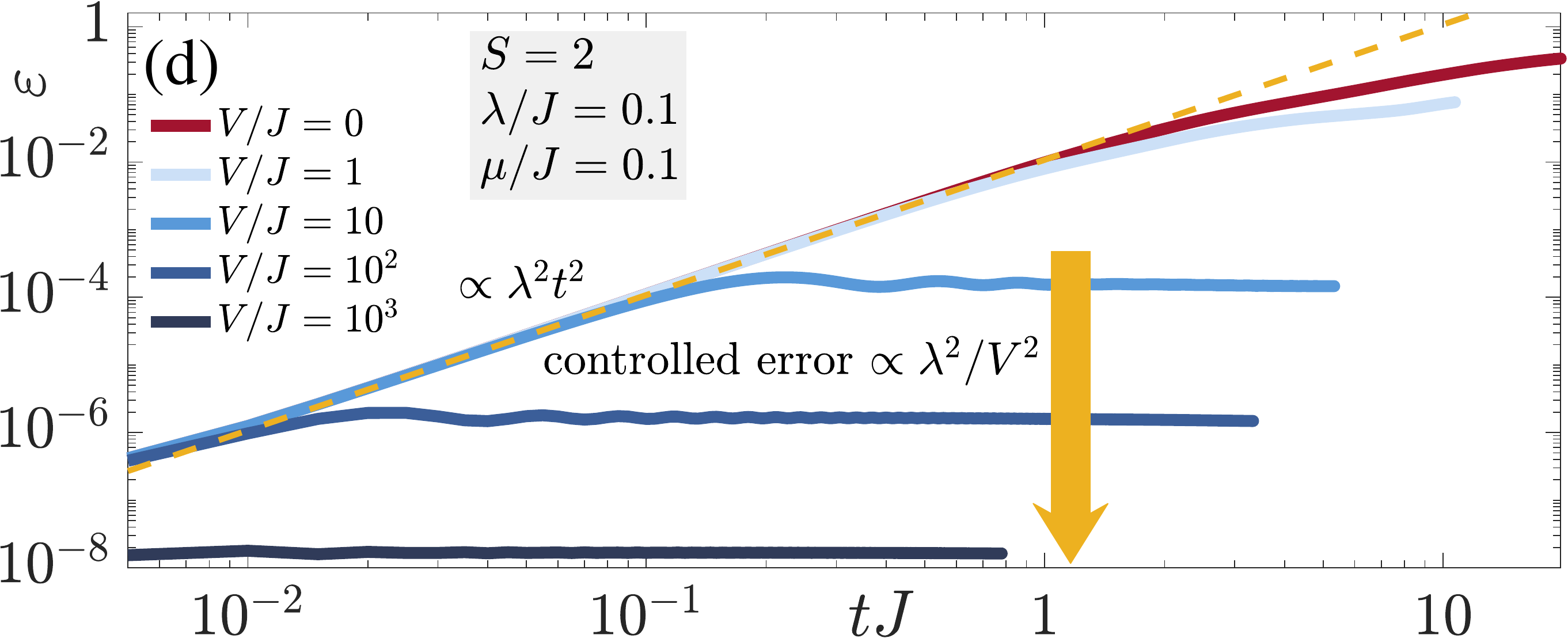}
	\caption{(Color online). Same as Fig.~\ref{fig:FullProtection} but with the energy-protection term given by the single-body Hamiltonian $V\tilde{H}_G=V\sum_jc_jG_j$ of Eq.~\eqref{eq:LinearProtection} with $c_j=(-1)^{j+1}$ rather than the full protection Hamiltonian $H_G$ of Eq.~\eqref{eq:FullProtection}. Even with a noncompliant sequence, the linear protection works remarkably well, and yields results qualitatively identical to those under full protection for the evolution times accessible in iMPS, including the gauge-violation plateau $\propto\lambda^2/V^2$ that occurs at the timescale $\propto1/V$.}
	\label{fig:LinearProtection} 
\end{figure*}

\begin{figure}[htp]
\centering
\includegraphics[width=.48\textwidth]{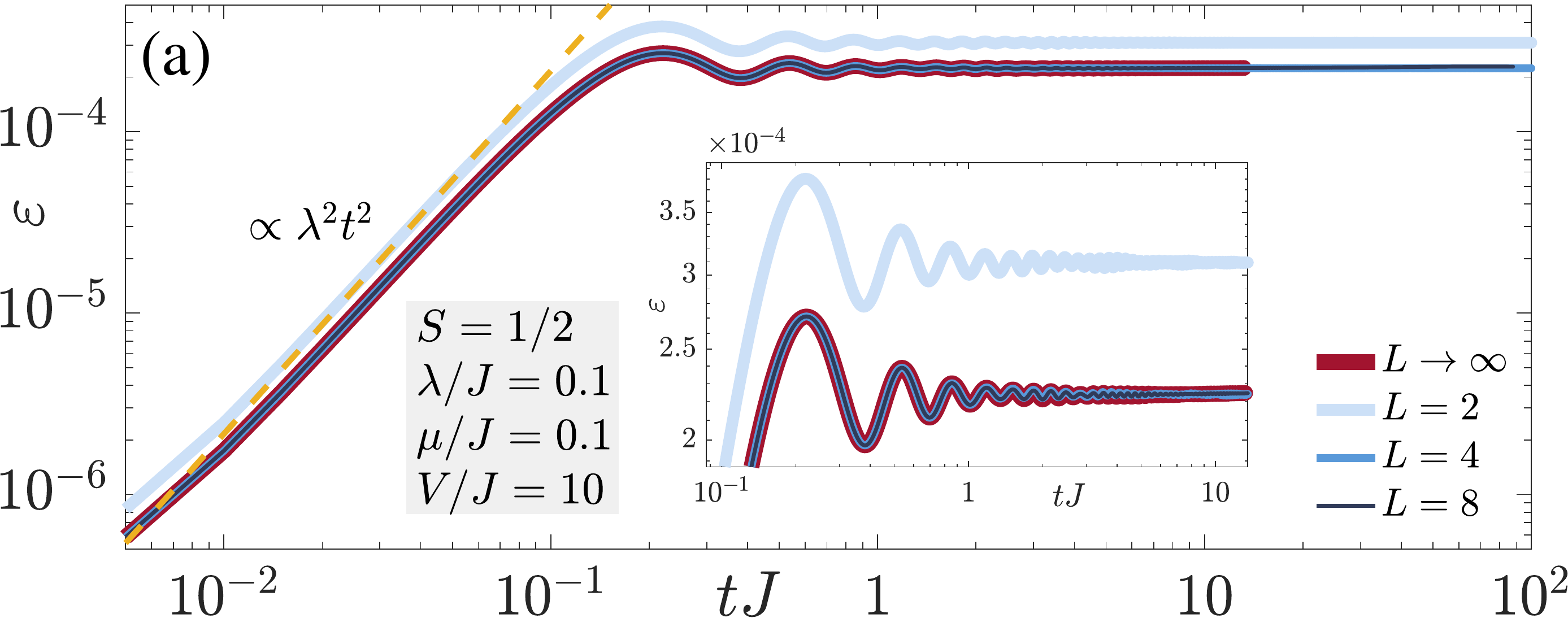}\\
\vspace{1.5mm}
\includegraphics[width=.48\textwidth]{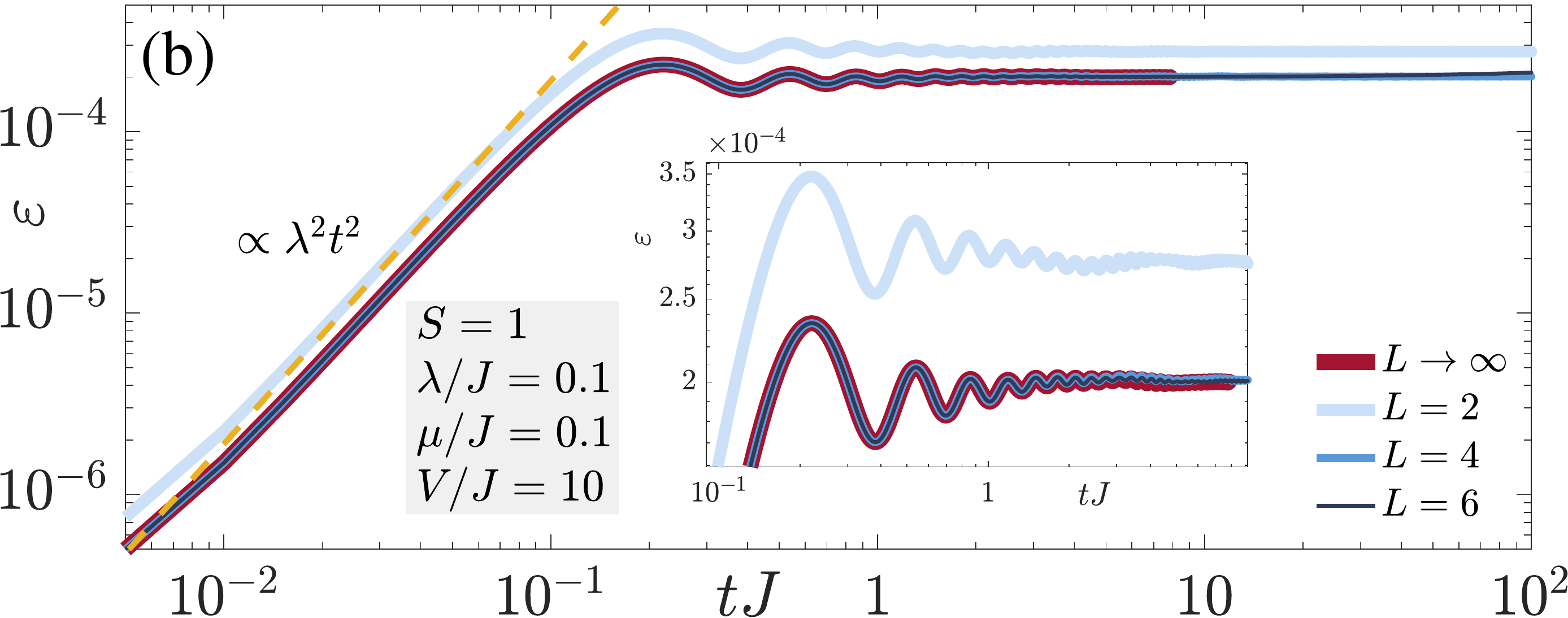}
\caption{(Color online). Same as Fig.~\ref{fig:FSSfull} but with the energy-protection term given by the single-body Hamiltonian $\tilde{H}_G$ of Eq.~\eqref{eq:LinearProtection} with $c_j=(-1)^{j+1}$ rather than the `full' protection Hamiltonian $H_G$ of Eq.~\eqref{eq:FullProtection}. Also in this case of linear protection with a noncompliant sequence, convergence to the thermodynamic limit is very fast over the evolution times accessible to iMPS.}
\label{fig:FSSlin}
\end{figure}

\subsection{Linear protection}\label{sec:lin}
Recently, it has also been shown  that reliable gauge protection can be achieved using single-body energy terms\cite{Halimeh2020e} of the form
\begin{align}\label{eq:LinearProtection}
V\tilde{H}_G=V\sum_jc_jG_j,
\end{align}
with properly chosen normalized rational \textit{compliant} coefficients $c_j\in[-1,1]$ that satisfy the condition $\sum_jc_jg_j=0$ if and only if $g_j=0,\,\forall j$. Here, the `local charges' $g_j$ are the eigenvalues of the generators $G_j$ of Gauss's law. 

The analysis from the constrained quantum dynamics discussed in Appendix~\ref{sec:constrained} does not apply here, since it requires that the protected target space be the ground state of $H_{\rm pro}$, which is not satisfied for the linear protection. In contrast, the analysis based on the quantum Zeno effect\cite{facchi2002quantum,facchi2004unification,facchi2009quantum,burgarth2019generalized} as discussed in Appendix~\ref{sec:QZE} can be adapted to this scenario. 
As derived within this formalism, linear protection with a compliant or even noncompliant sequence leads also to an adjusted gauge theory $H_0+\lambda\mathcal{P}_0H_1\mathcal{P}_0$ for local observables up to a system size-dependent error 
\begin{align}
	\big\lvert\langle e^{iHt}Oe^{-iHt} - e^{iH_\text{adj}t}Oe^{-iH_\text{adj}t}\rangle\big\rvert \leq \tilde{\Delta}_\text{adj},
\end{align}
with $\tilde{\Delta}_\text{adj}\sim tV_0^2L^2/V$. The specific form of  $\tilde{\Delta}_\text{adj}$ is given in Appendix~\ref{sec:QZE}. This bound is therefore ensured up to the timescale $\tilde{\tau}_\text{adj}\propto V/(V_0L)^2$. A renormalized gauge theory dominates at later times, but only for a compliant sequence, and persists up to an exponential timescale $\tau_\text{ren}\propto\exp(V/V_0)/V_0$, which is similar to its counterpart in the case of full protection; cf.~Appendix~\ref{sec:Abanin}. Unlike $\tilde{\tau}_\text{adj}$, the timescale $\tau_\text{ren}$ does not explicitly depend on the system size $L$, but the reliability of dynamics nevertheless does. As mentioned above, the compliant sequence $c_j$ is a normalized set of rational numbers, such that the largest $|c_j|$ is unity: As $L$ increases, this means that the spacing between the different $c_j$ will get smaller on average. For a fixed value of $\lambda$, this in turn implies that $V$ must be increased to achieve the same level of  provable reliability in the gauge invariance up to a given evolution time.

The compliance of $c_j$ is a sufficient albeit not a necessary condition when the gauge-breaking errors are local. Indeed, in Ref.~\onlinecite{Halimeh2020e} it is demonstrated numerically for finite systems that for specific error terms even \textit{noncompliant} sequences such as $c_j=(-1)^{j+1}$ can still suppress the gauge violation $\propto\lambda^2/V^2$ up to indefinite times. Furthermore, such a noncompliant sequence can be shown\cite{facchi2002quantum,facchi2004unification,facchi2009quantum,burgarth2019generalized,Halimeh2020e} to still give rise to at least the adjusted gauge theory $H_0+\lambda\mathcal{P}_0H_1\mathcal{P}_0$ up to the timescale $\tilde{\tau}_\text{adj}\propto V/(V_0L)^2$, just like its compliant counterpart, albeit no analytic proof of a renormalized gauge theory up to a timescale exponential in $V$ exists.

A strong and essential feature of iMPS is translation invariance, and as such a compliant sequence is not optimal in an iMPS implementation as it breaks this symmetry. In contrast, a sequence such as $c_j=(-1)^{j+1}$ preserves translation invariance down to a unit cell of two adjacent matter sites. Using this feature, we now quench the target-sector initial state $\ket{\psi_0}$ of Fig.~\ref{fig:illustration} by the Hamiltonian $H=H_0+\lambda H_1+V\tilde{H}_G$, with $\tilde{H}_G=\sum_j(-1)^{j+1}G_j$. The resulting dynamics of the gauge violation is plotted in Fig.~\ref{fig:LinearProtection} for various values of the link spin length $S$. Qualitatively, the results are identical to those of Fig.~\ref{fig:FullProtection}. The gauge violation initially grows $\propto\lambda^2t^2$, in agreement with TDPT, before plateauing at a timescale $\propto1/V$ upon which the protection term dominates. The plateau value at sufficiently large $V$ is $\propto\lambda^2/V^2$, as expected from degenerate perturbation theory.\cite{Halimeh2020a} As in the case of full protection, linear protection affords reliable gauge invariance at any link spin length $S$ within the accessible evolution times in iMPS, even with the noncompliant sequence $c_j=(-1)^{j+1}$. This is remarkable as it shows that even in the thermodynamic limit there is an experimentally feasible single-body protection scheme with the simple coefficients $\pm1$ that allows for well-controlled gauge-theory dynamics at least up to times that are relevant for current experiments. It is interesting to note that we do not find a deterioration of the reliability of gauge invariance when increasing the link spin length $S$. On the contrary, with our particular experimentally relevant error given in Eq.~\eqref{eq:error}, gauge invariance seems to become even more reliable with larger $S$ for a fixed value of $\lambda$ and $V$.

\begin{figure*}[htp]
	\centering
	\includegraphics[width=.48\textwidth]{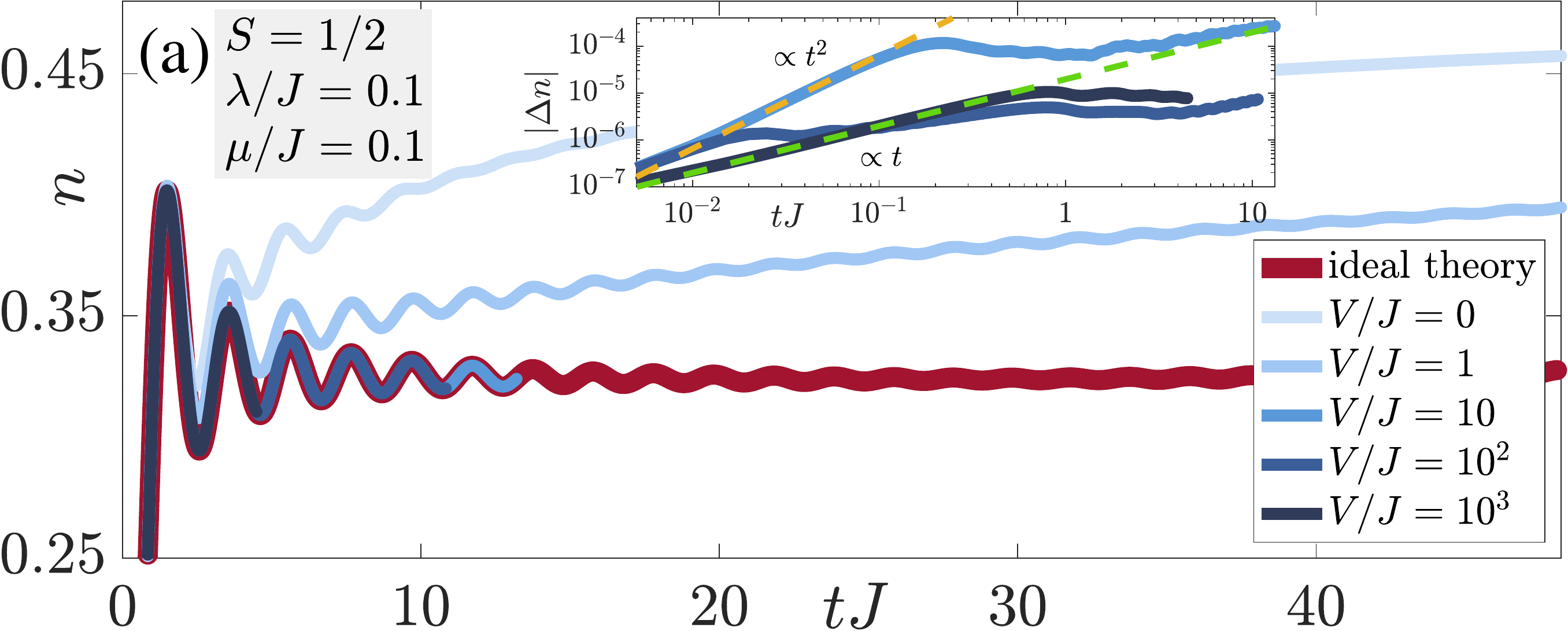}\quad
	\includegraphics[width=.48\textwidth]{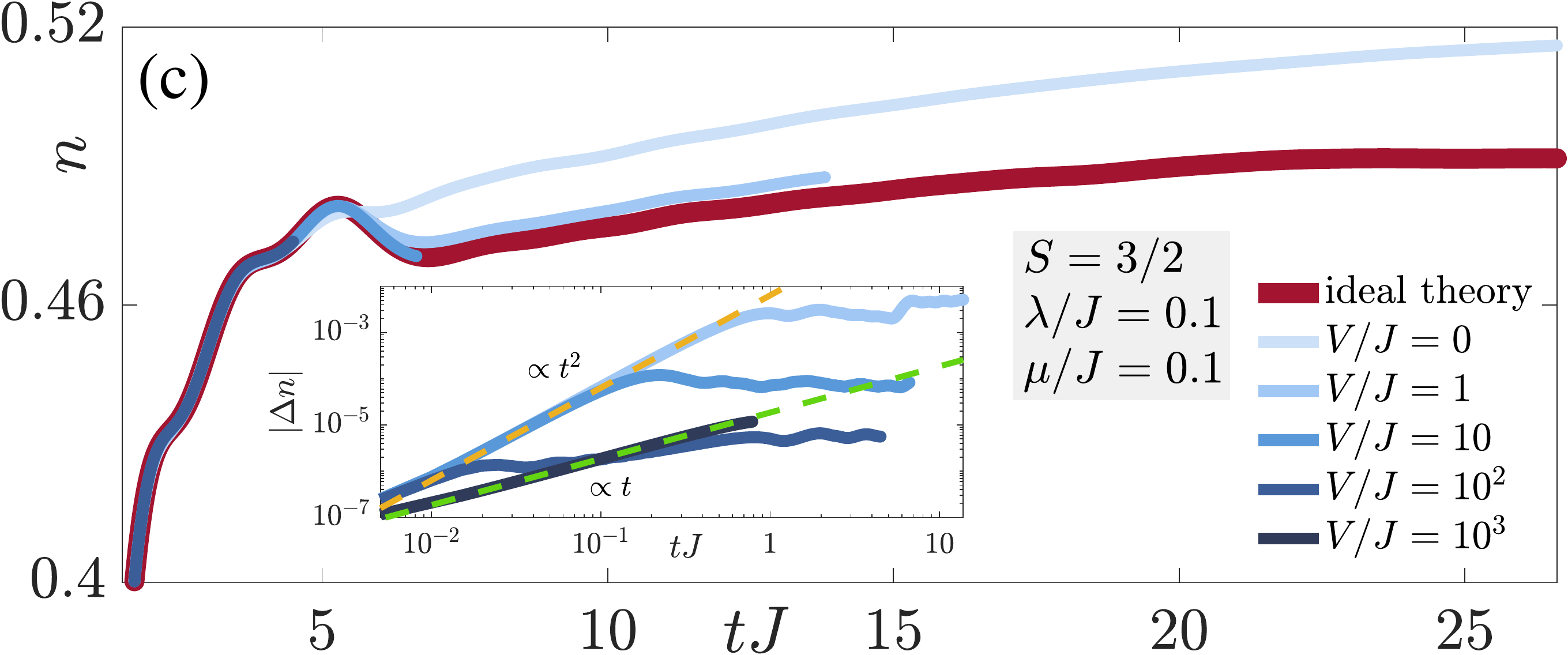} \\
	\includegraphics[width=.48\textwidth]{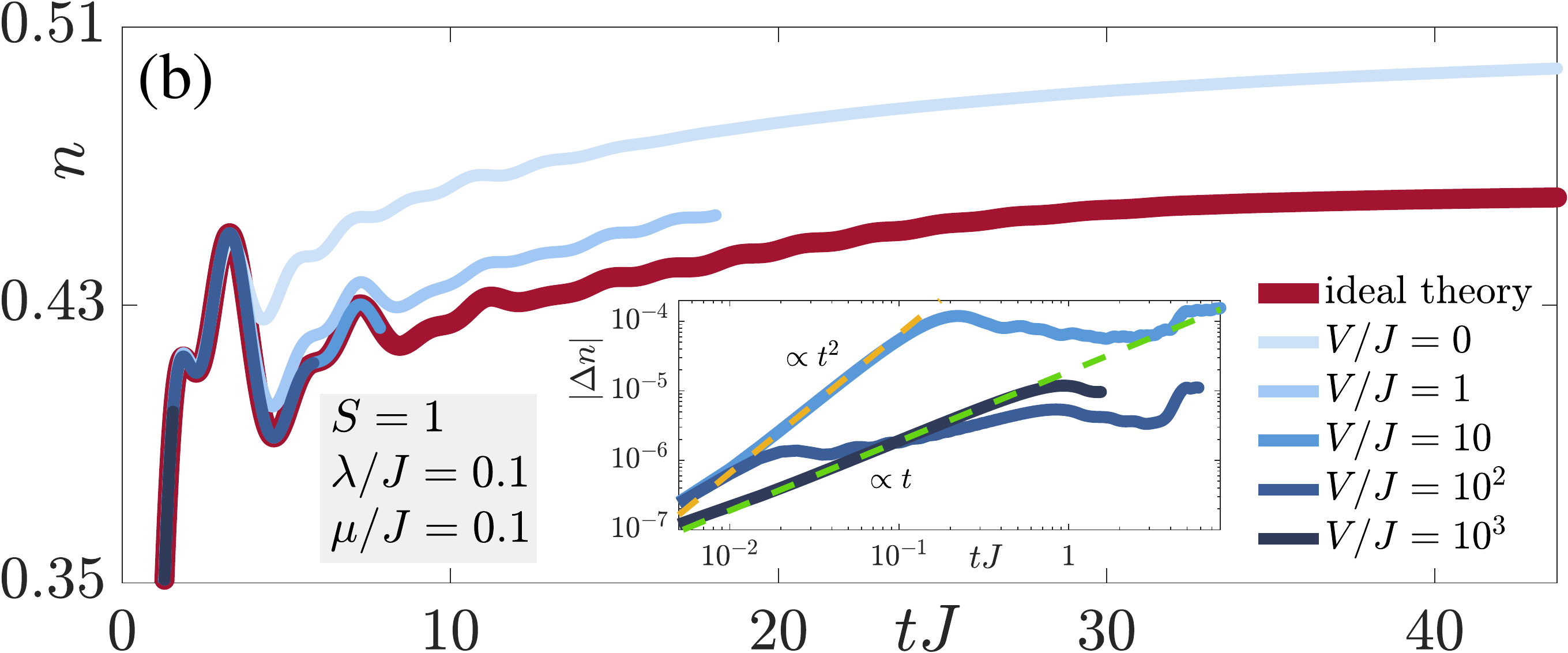}\quad
	\includegraphics[width=.48\textwidth]{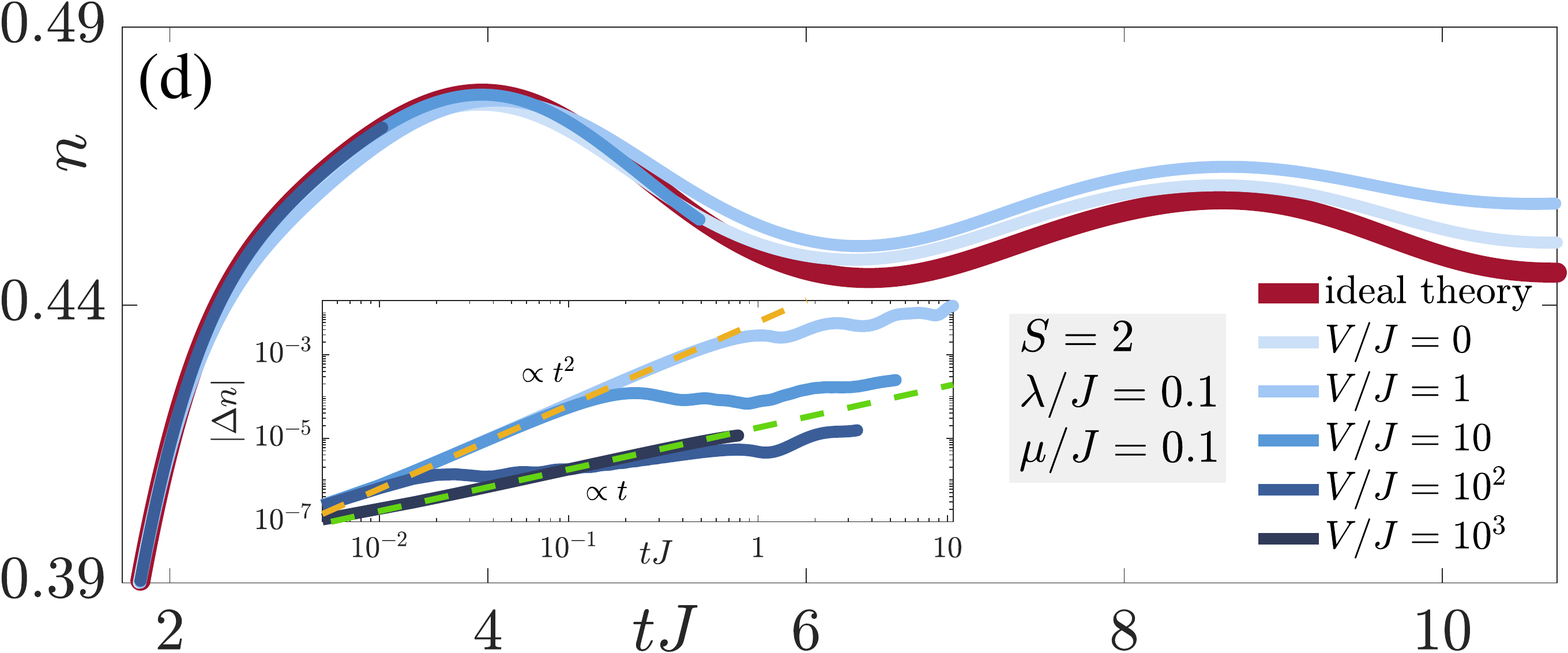}
	\caption{(Color online). Same as Fig.~\ref{fig:FullProtection_sigmaz} but using the linear-protection term with a noncompliant sequence, $V\tilde{H}_G=V\sum_j(-1)^{j+1}G_j$, rather than full protection. The dynamics under the faulty gauge theory $H=H_0+\lambda H_1+V\tilde{H}_G$ reliably approaches the ideal-theory dynamics under $H_0$ (which is also the adjusted gauge theory in the case of our gauge-breaking term) for $V\gtrsim10$. The insets show the deviation $\lvert\Delta n\rvert$ of the faulty dynamics from its ideal counterpart, where the error initially grows at $\propto t^2$, but then slower ($\propto t$) at later times, with this transition occurring at a timescale $\propto1/V$ at which the protection term begins to dominate. This is milder than the upper error bound with respect to the adjusted gauge theory (see main text and Appendix~\ref{sec:QZE}).}
	\label{fig:LinProtection_sigmaz} 
\end{figure*}
\begin{figure}[htp]
\centering
\includegraphics[width=.48\textwidth]{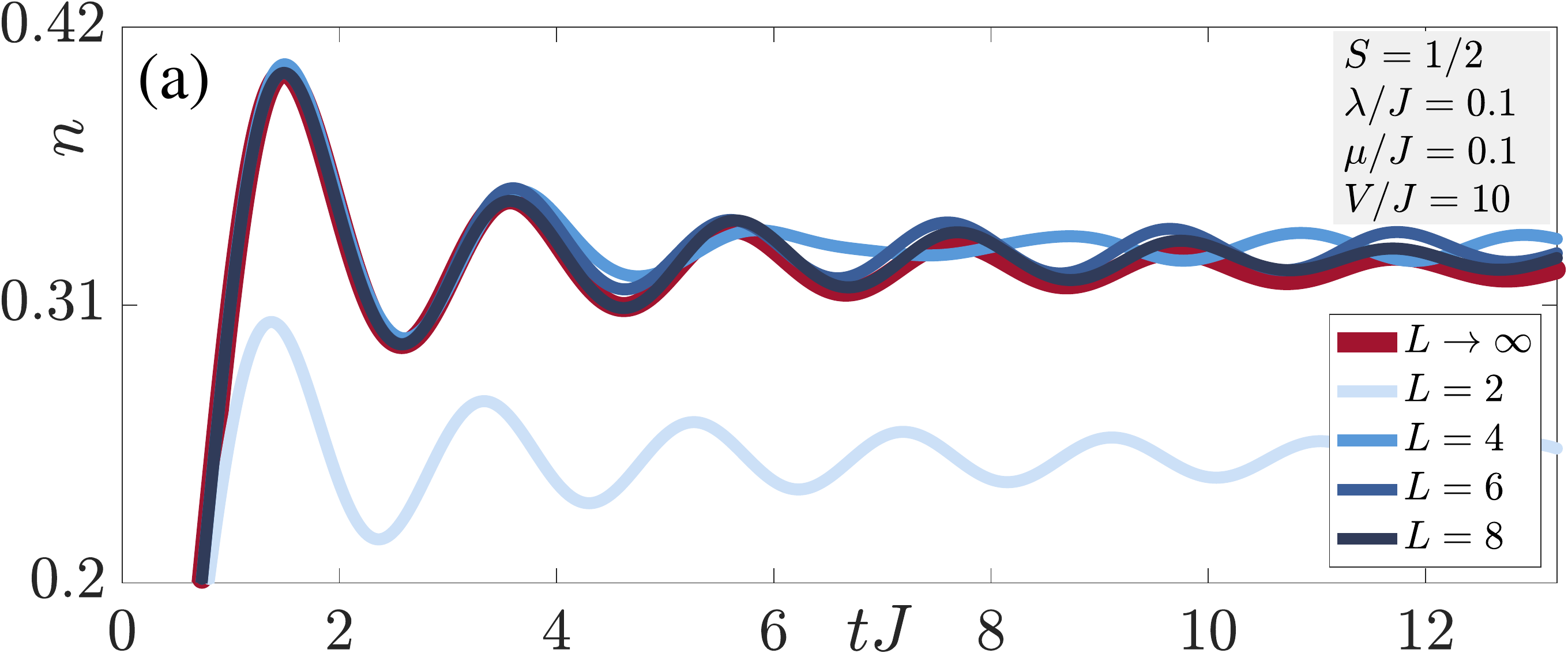}\\
\vspace{1.5mm}
\includegraphics[width=.48\textwidth]{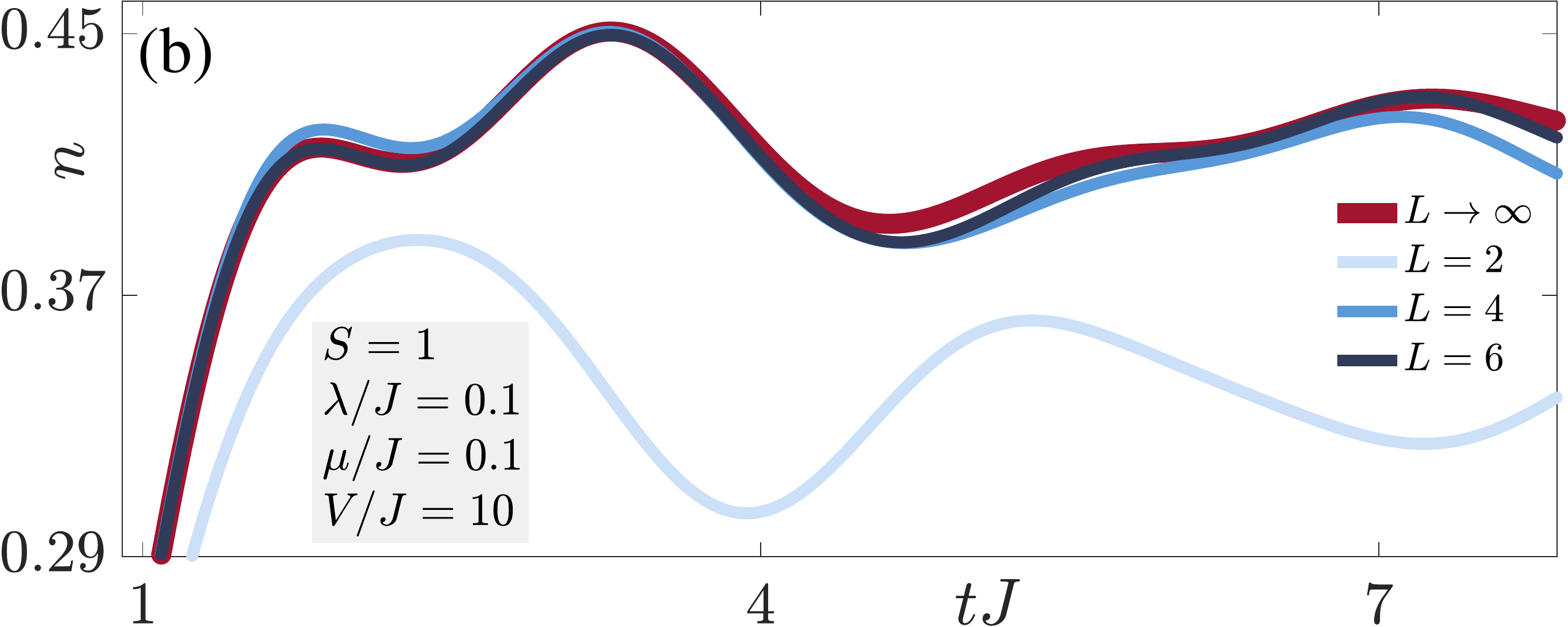}
\caption{(Color online). Same as Fig.~\ref{fig:FSSfull_sigmaz}, but where gauge invariance is supported through the linear-protection term with a noncompliant sequence, $V\tilde{H}_G=V\sum_j(-1)^{j+1}G_j$. Also in this case, convergence to the thermodynamic limit occurs at just a few matter sites during the evolution times accessible in iMPS.}
\label{fig:FSSlin_sigmaz} 
\end{figure}

We now analyze the finite-size behavior of the gauge violation in the case of the linear protection term $V\tilde{H}_G=V\sum_j(-1)^{j+1}G_j$, as shown in Fig.~\ref{fig:FSSlin}(a,b) for link spin lengths $S=1/2$ and $1$, respectively. The same qualitative conclusions from the full-protection case (see Fig.~\ref{fig:FSSfull_sigmaz}) hold in that the dynamics of finite systems calculated through ED quickly converge to the dynamics in the thermodynamic limit, which is calculated in iMPS. Indeed, for both spin lengths considered, already the result for $L=4$ matter sites almost coincide with that of the thermodynamic limit. This behavior is intriguing because in the case of linear protection the system-size dependence of the timescale $\tilde{\tau}_\text{adj}\propto V/(JL)^2$ for the adjusted gauge theory may lead one to expect 
a slow approach to the thermodynamic limit $L\to\infty$. Our numerical results indicate that, quite to the contrary, this approach is as fast as in the case of full protection, at least up to the accessible times in iMPS and for the local error presented in the article.

It is instructive to also study the performance of linear protection with a noncompliant sequence for local observables such as the particle density. The results are shown in Fig.~\ref{fig:LinProtection_sigmaz} for various values of the link spin length $S$, where we compare the time evolution of the particle density upon quenching $\ket{\psi_0}$ with $H=H_0+\lambda H_1+V\tilde{H}_G$ to that after quenching $\ket{\psi_0}$ with the adjusted gauge theory $H_\text{adj}=H_0+\lambda\mathcal{P}_0H_1\mathcal{P}_0=H_0$. As in the case of full protection, the protected dynamics quickly approaches its ideal-theory counterpart with larger $V$ for all values of $S$ considered. In the insets, we also see an initial growth $\propto t^2$ in the deviation $\lvert\Delta n\rvert$ that transitions to a growth $\propto t$ at later times for a fixed $V$. The time of this transition decreases with $V$ just as in the case of full protection. Actually, it seems that this transition timescale $\propto1/V$, upon which the protection term begins to dominate. However, in contrast to the case of full protection, we do not find a suppression by $V$ in the deviation. Indeed, the adjusted gauge theory is valid up to an error upper bound $\propto tV_0^2L^2/V$, which is $L$-dependent, unlike in the case of full protection. How precisely the error grows within this bound may therefore be influenced by the system size in different ways depending on the value of $V$ and on the observable considered. In the case of full protection (see Sec.~\ref{sec:full} and Appendix~\ref{sec:Abanin}), where the error upper bound $\propto t^2V_0^3/V$ is fully independent of system size, $L$ does not seem to play a role in the deviation growth at early times. Nevertheless, it is remarkable that up to even intermediate evolution times the dynamics under the faulty gauge theory $H$ is still in excellent agreement with that of the adjusted gauge theory, as shown in Fig.~\ref{fig:LinProtection_sigmaz}. This behavior far exceeds the expectations laid out in our analytics, and suggests that experimentally relevant local errors are quite reliably controlled even under linear protection with noncompliant sequences.

Finally, we study the finite-size behavior of the particle density in Fig.~\ref{fig:FSSlin_sigmaz}(a,b) for link spin lengths $S=1/2$ and $1$, respectively. Qualitatively similar to the case of full protection, the dynamics calculated in ED for finite $L$ quickly approach the dynamics in the thermodynamic limit $L\to\infty$ calculated in iMPS. So not only does linear protection with a simple noncompliant sequence $c_j=(-1)^{j+1}$ provide reliable gauge invariance up to experimentally relevant times, it also faithfully reproduces observable dynamics even in the thermodynamic limit. This is despite the fact that our analytic derivations of a volume-dependent timescale $\tilde{\tau}_\text{adj}\propto V/(V_0L)^2$ in the case of linear protection predict that agreement with the adjusted gauge theory ($H_0$ in our case) may not happen in the thermodynamic limit in a worst-case scenario. Once again, this affirms the suggestion already extracted from Fig.~\ref{fig:LinProtection_sigmaz} that local errors such as those of Eq.~\eqref{eq:error} are faithfully controlled by linear protection even with a simple sequence $c_j=(-1)^{j+1}$. In Appendix~\ref{sec:electric_lin}, we provide similar results for the electric field, which support the above conclusions.

For nonlocal error terms the noncompliant sequence is in general not sufficient.\cite{Halimeh2020e} However, typical experimental errors have a large degree of locality, so we expect to obtain similarly good performance in most relevant situations, potentially by slightly increasing the unit cell of the noncompliant sequence.

\section{Summary of emergent gauge theories and timescales}\label{sec:timescale}
We summarize here all the timescales and emergent gauge theories discussed in Sec.~\ref{sec:quench}, and their dependence on the kind of protection used (i.e., full or linear). 

\subsection{Adjusted gauge theory}
Independent of whether full protection $VH_G=V\sum_jG_j^2$ or linear protection $V\tilde{H}_G=V\sum_jc_jG_j$ (with a compliant or noncompliant sequence $c_j$) is used to protect against gauge-breaking errors $H_1$, an \textit{adjusted} gauge theory $H_\text{adj}=H_0+\lambda\mathcal{P}_0H_1\mathcal{P}_0$ emerges. It is valid up to a given timescale, during which the error in the dynamics of a local observable is controllably bounded from above with respect to the adjusted gauge theory. In the case of full protection, this upper bound is $\propto t^2V_0^3/V$, leading to the timescale of $\tau_\text{adj}\propto\sqrt{V/V_0^3}$. Derivational details are provided in Appendix~\ref{sec:constrained}. In the case of linear protection, the upper bound is $\propto tV_0^2L^2V$, yielding the timescale $\tilde{\tau}_\text{adj}\propto V/(V_0L)^2$. Derivational details are provided in Appendix~\ref{sec:QZE}. In both cases, the energy term $V_0$ is roughly a linear sum of $\{\lambda,g^2aS^2,\mu,J\}$, and its exact form is given in Eq.~\eqref{eq:V0} in Appendix~\ref{sec:Abanin}.

As such, we find that the dynamics is related to an adjusted gauge theory up to a fractional (in $V$) timescale that is volume-independent in the case of full protection, but which depends on system size in the case of linear protection. However, our numerical results in Sec.~\ref{sec:lin} indicate that system size can play a trivial role in the case of linear protection even when the sequence $c_j$ is noncompliant. This is evidenced by the excellent agreement of the dynamics of the particle density under the faulty theory $H=H_0+\lambda H_1+V\sum_j(-1)^{j+1}G_j$ and that under the adjusted theory in the thermodynamic limit, as shown in Fig.~\ref{fig:LinProtection_sigmaz}. Furthermore, the finite-size behavior of the particle density in Fig.~\ref{fig:FSSlin_sigmaz} also indicates that convergence to the thermodynamic limit is fast in this case (results for the electric field in Appendix~\ref{sec:electric_lin} yield the same conclusion). We can therefore conclude that in the case of experimentally relevant local gauge-breaking terms as those of Eq.~\eqref{eq:error}, even linear protection with a simple noncompliant sequence $c_j=(-1)^{j+1}$ can reliably reproduce the adjusted gauge theory dynamics in the thermodynamic limit.

\subsection{Renormalized gauge theory}
In the case of full protection or linear protection with a \textit{compliant} sequence, a renormalized gauge theory emerges lasting up to an exponential timescale $\tau_\text{ren}\propto\exp(V/V_0)/V_0$. The derivation of this renormalized gauge theory is based on the ARHH framework,\cite{abanin2017rigorous} and has recently been adapted to gauge theories in Ref.~\onlinecite{Halimeh2020e}. Even though $\tau_\text{ren}$ is not explicitly volume-dependent, a given level of gauge invariance at a fixed error strength $\lambda$ will require $V$ to increase with system size in the case of linear protection with a compliant sequence but not in the case of full protection. Derivational details of the renormalized gauge theory is presented in Appendix~\ref{sec:Abanin}. The same degree of rigorously provable reliability cannot be achieved for linear protection with noncompliant sequences, as in that case gauge breakings to high order could become resonant and thus spoil the gauge symmetry at fractionally large times (although our numerical results suggest a very favorable behavior even in this case).

\section{Conclusion}\label{sec:conc}
We have presented infinite matrix product state calculations for quench dynamics in the paradigmatic spin-$S$ $\mathrm{U}(1)$ quantum link model in the presence of experimentally relevant local gauge-breaking errors, against which we protect with either full or linear protection terms. We have found that reliable gauge invariance is achieved in the thermodynamic limit for all accessible evolution times. 

In the case of full protection, our numerical results support our analytic predictions that quench dynamics under a faulty gauge theory is reproduced by an \textit{adjusted} gauge theory. Analytic calculations demonstrate this adjusted gauge theory to hold up to a volume-independent error bound that grows proportional to the inverse of  the protection strength. Remarkably, our numerical results show that for the local gauge-breaking terms considered the actual deviation of local observables grows well below this bound. This allows the dynamics under the faulty theory at sufficiently large protection strength to agree very well with that under the adjusted gauge theory for all times accessible to our simulations. Exact diagonalization results at just a few matter sites converge to those in the thermodynamic limit calculated through infinite matrix product state techniques, indicating a wide-reaching volume-independence of full protection.

In the case of linear protection with a noncompliant sequence, an adjusted gauge theory also emerges, though analytics predicts an error bound that is inversely proportional to the protection strength and grows as the square of system size. Nevertheless, our numerical results indicate a much milder behavior even in the thermodynamic limit, with the dynamics under the faulty theory at sufficiently large protection strength agreeing very well with that of the adjusted gauge theory up to all accessible evolution times. In fact, we see that linear protection with a simple alternating noncompliant sequence $\pm1$ fares qualitatively just as well as full protection. Additionally, we have also shown through exact diagonalization that the dynamics converges quickly to the thermodynamic limit, even at just a few matter sites, regardless of the link spin length. This bodes well for experimental efforts attempting to realize the QED limit of lattice gauge theories on intermediate-scale quantum devices, because our findings support the conclusion that for experimentally relevant local gauge-breaking errors, reliable gauge invariance can be achieved in the thermodynamic limit for large link spin length throughout evolution times typical of modern setups. Even more, depending on the error form, this can be possible with linear-protection sequences as simple as a local chemical potential that alternates as $\pm1$ on odd and even matter sites.

It is worth mentioning that in this work we have only considered a $(1+1)-$D Abelian gauge theory. It would be interesting to extend these studies to non-Abelian gauge theories where local generators do not commute. Nevertheless, investigations on $(1+1)-$D Abelian gauge theories are highly relevant for ongoing experiments.\cite{Goerg2019,Schweizer2019,Mil2020,Yang2020}
Moreover, the conclusions of our work, namely that gauge invariance in the quench dynamics of a \textit{faulty} gauge theory can be reliably achieved in the thermodynamic limit through energy protection, complement the findings of Ref.~\onlinecite{VanDamme2020} in equilibrium for the same $(1+1)-$D Abelian gauge theory (see also Ref.~\onlinecite{Borla2020} for similar conclusion in equilibrium for a \textit{gauged Kitaev chain}). There, full protection of sufficiently large strength produces a renormalized gauge theory. This behavior is possible because the full protection term lends mass to the Higgs boson with which the gauge-breaking errors couple.\cite{Poppitz2008,Kuno2015,Bazavov2015} Another current frontier for gauge quantum simulations is presented by higher dimensions.\cite{ Paulson2020simulating,Ott2020scalable} The resilience and protection against gauge breaking errors in quantum link models and other lattice gauge theories in $(2+1)-$D and $(3+1)-$D, where a massless photon exists, is thus another interesting avenue for the future studies.

\section*{Acknowledgments}
MVD and HL contributed equally to this work. JCH is grateful to Fred Jendrzejewski, Guoxian Su, and Bing Yang for discussions related to experimental aspects of this work. We acknowledge support by Provincia Autonoma di Trento, the ERC Starting Grant StrEnQTh (Project-ID 804305), Q@TN --- Quantum Science and Technology in Trento, ERC grants No 715861 (ERQUAF) and 647905 (QUTE), and from Research Foundation Flanders (FWO) via grant GOE1520N.

\appendix
\section{Supporting numerical results}\label{sec:support}

\begin{figure*}[htp]
	\centering
	\includegraphics[width=.48\textwidth]{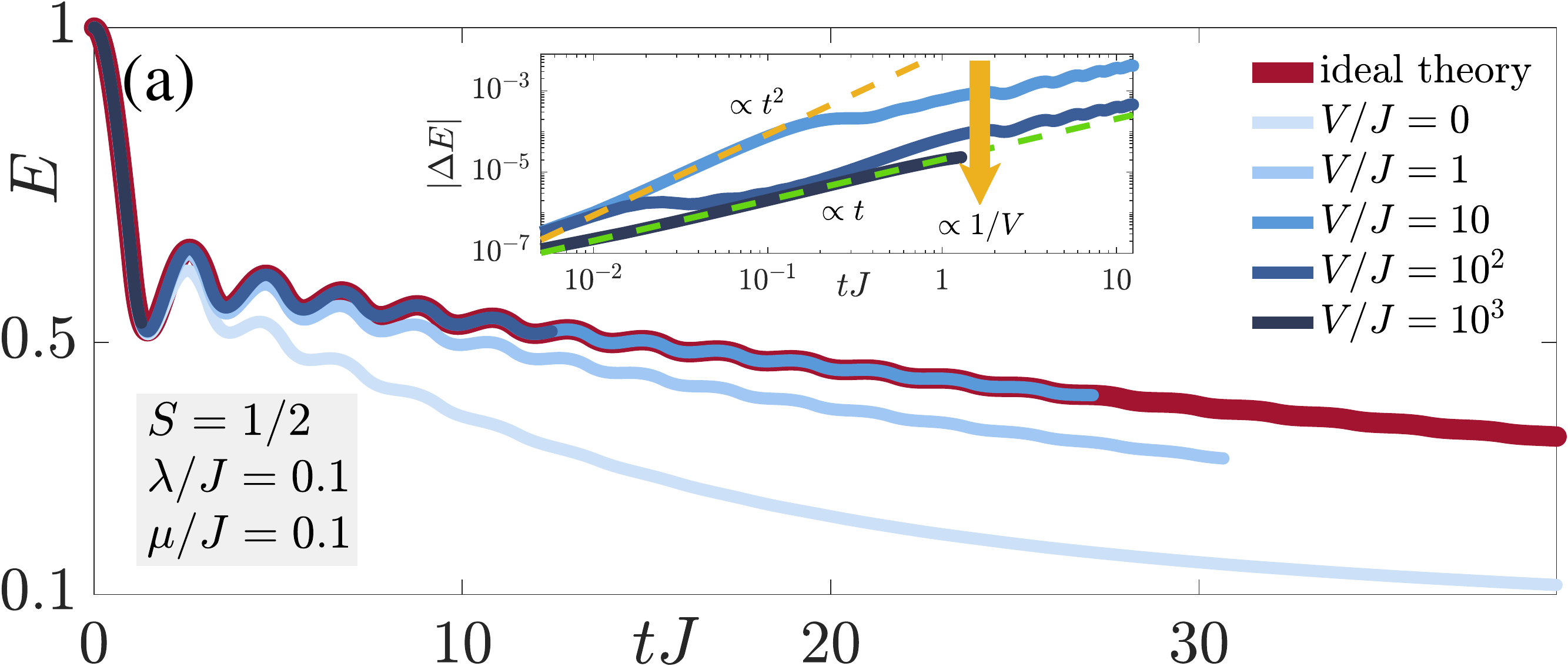}\quad
	\includegraphics[width=.48\textwidth]{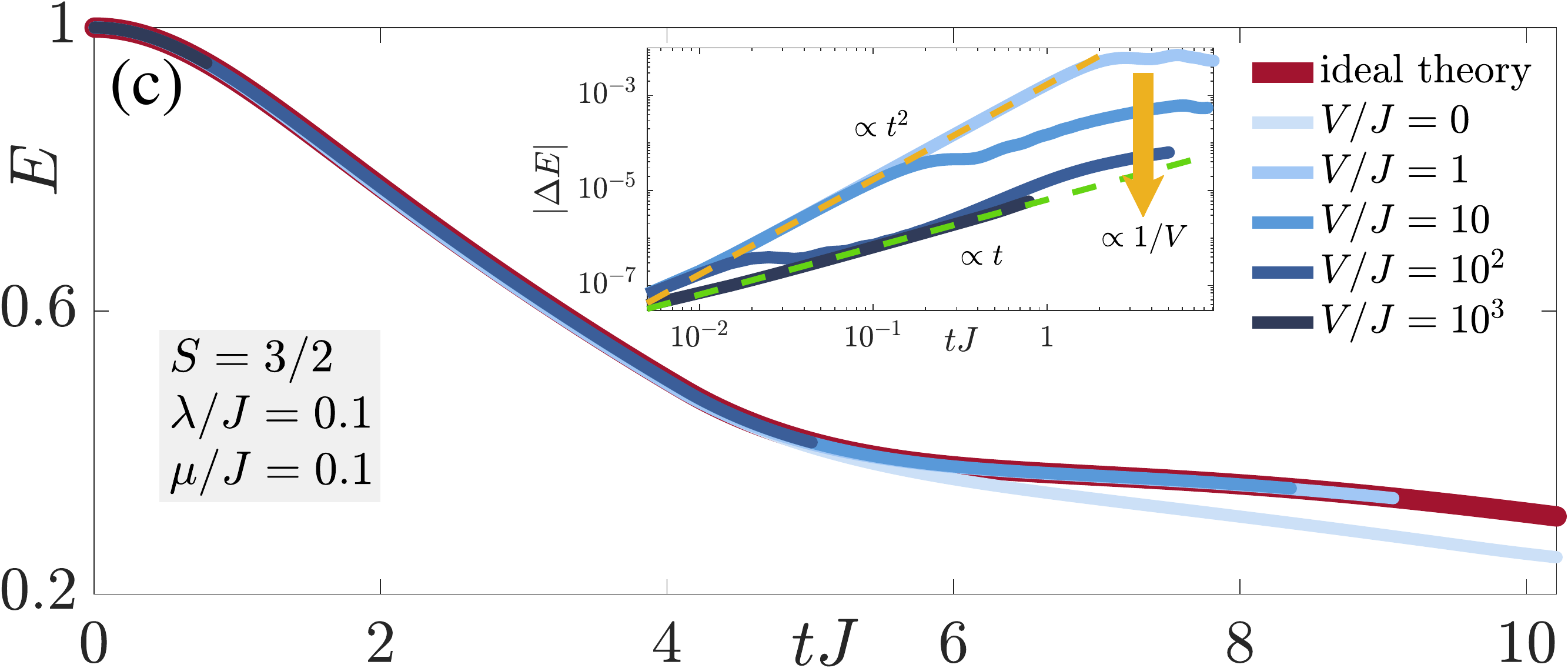} \\
	\includegraphics[width=.48\textwidth]{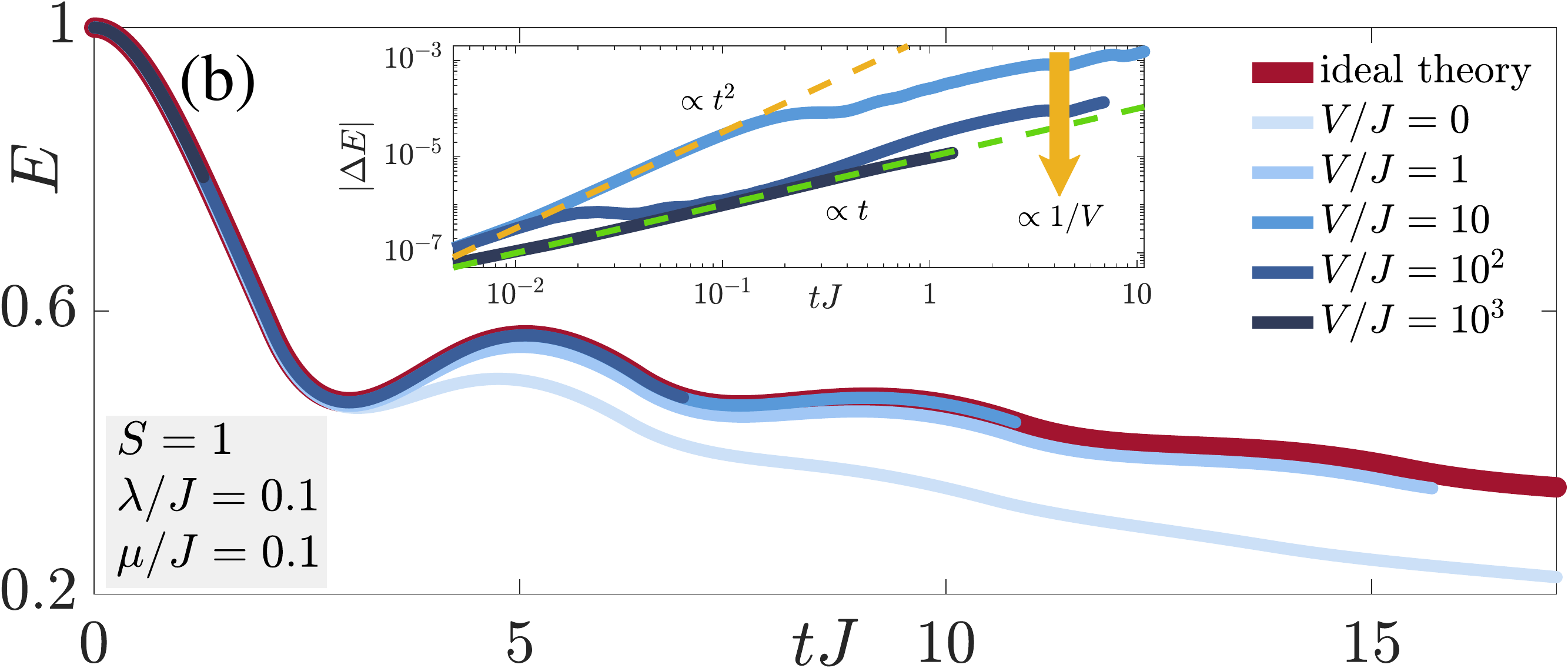}\quad
	\includegraphics[width=.48\textwidth]{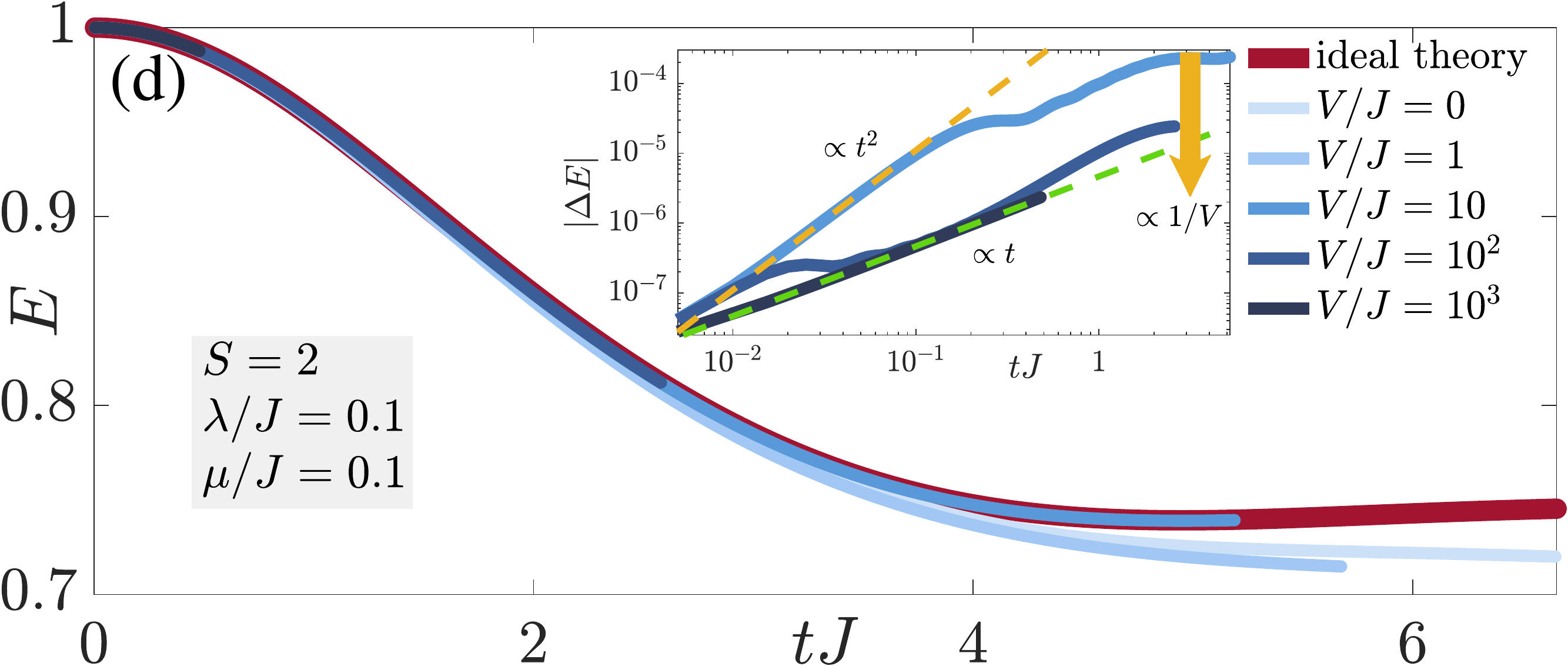}
	\caption{(Color online). Deviation of electric field from ideal dynamics due to gauge-breaking terms. Plotted is the temporally averaged absolute value of the electric field (normalized by spin length $S$) for gauge-breaking strength $\lambda/J=0$ (ideal dynamics; red line) and $\lambda/J=0.1$ with full protection $VH_G=V\sum_jG_j^2$ at various strengths $V$ (faulty gauge theory; shades of blue) for link spin lengths (a) $S=1/2$, (b) $S=1$, (c) $S=3/2$, and (d) $S=2$. The deviation from the ideal theory is rapidly suppressed as $V$ increases, at least up to the evolution times accessible by iMPS. With increasing $V$, the short-time scaling diminishes from $\propto t^2$ (the theoretical upper bound) to a milder scaling $\propto t$ at a timescale $\propto1/V$, after which the protection term begins to dominate.
	}
	\label{fig:FullProtection_efield} 
\end{figure*}

\begin{figure}[htp]
	\centering
	\includegraphics[width=.48\textwidth]{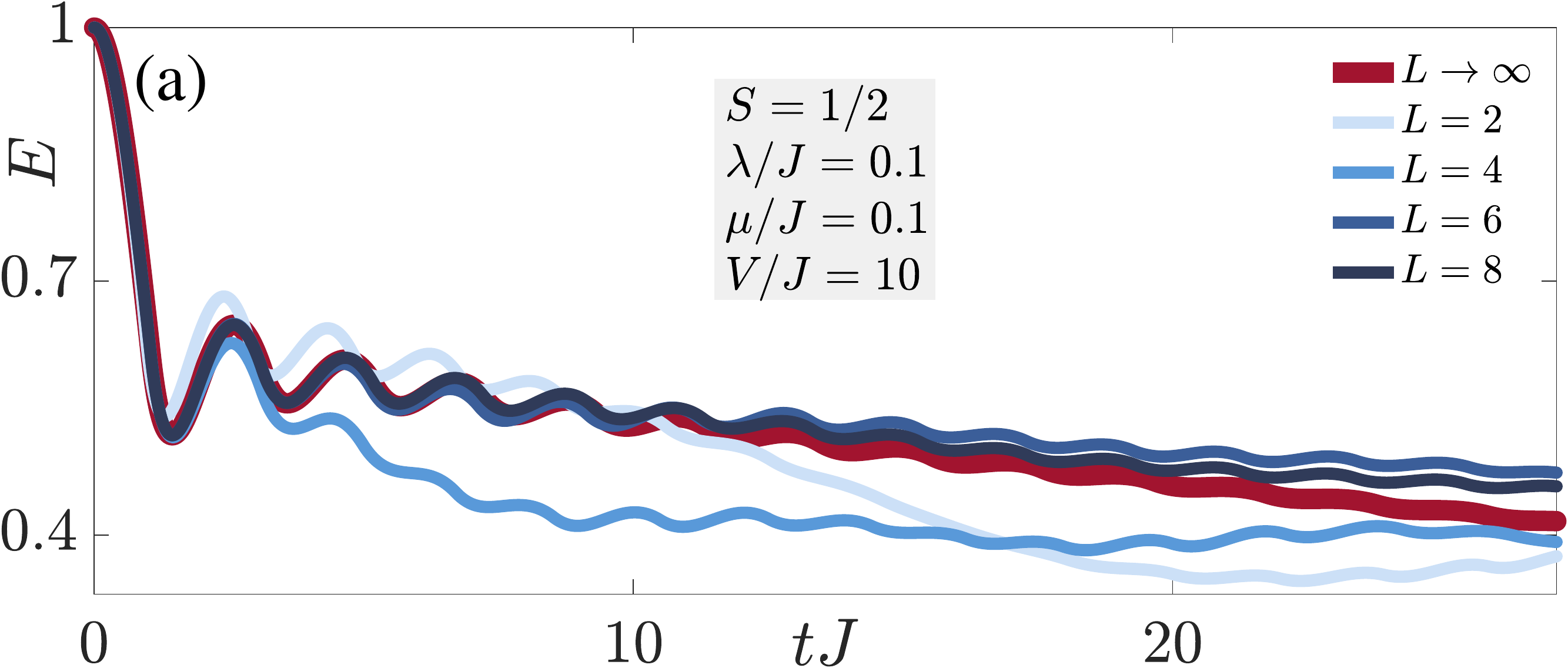}\\
	\includegraphics[width=.48\textwidth]{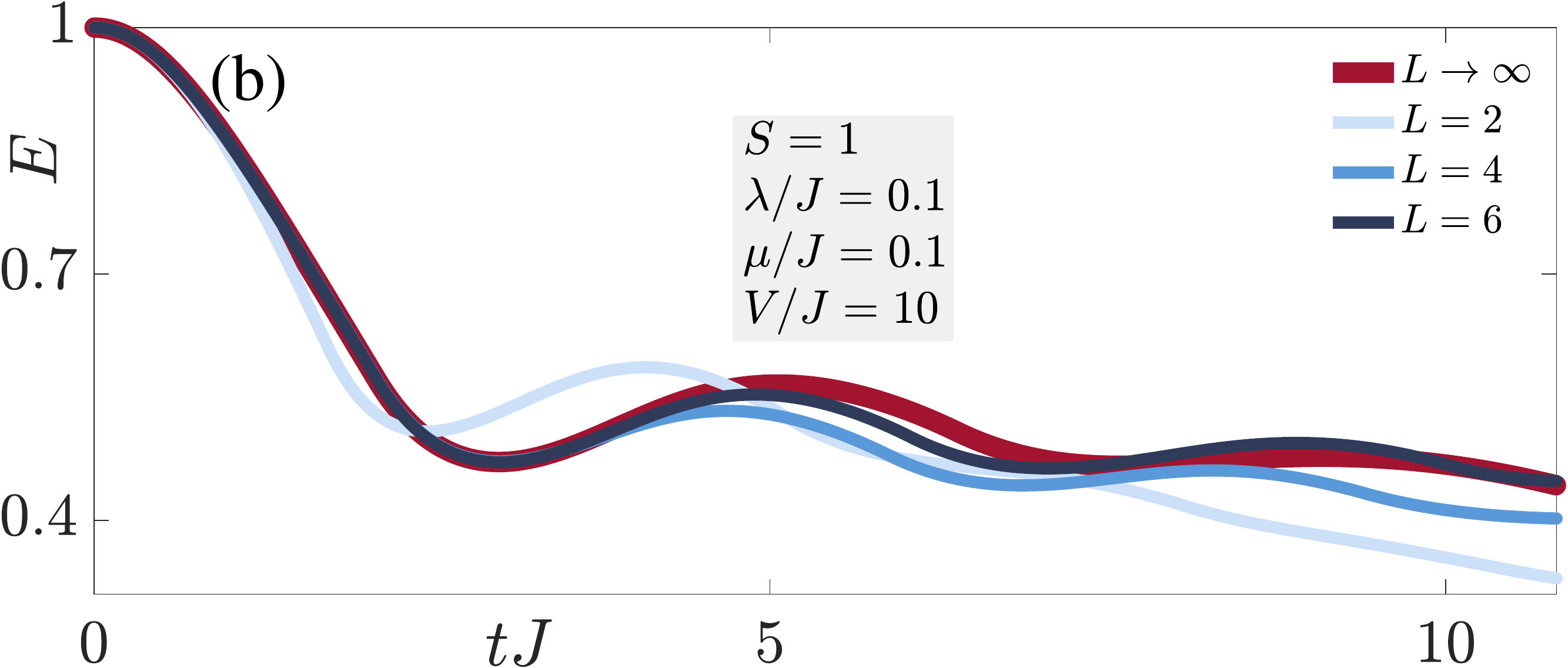}
	\caption{(Color online). Finite-size behavior of the electric field under full gauge protection. Parameters are $\lambda=0.1J$, $V=10J$, and quantum link spin length (a) $S=1/2$ and (b) $S=1$. As in the case of the gauge violation in Fig.~\ref{fig:FSSfull} and particle density in Fig.~\ref{fig:FSSfull_sigmaz}, the result quickly approaches its value in the thermodynamic limit already at a few matter sites.}
	\label{fig:FSSfull_efield} 
\end{figure}
\subsection{Results for electric field}\label{sec:electric}
In the main text, we have provided numerical results in both ED and iMPS for the gauge violation and particle density in the case of both full and linear protection. Here, we supplement these results by showing their counterparts for the electric field normalized by the link spin length $S$, as given in Eq.~\eqref{eq:electric}. Conclusions are qualitatively the same as those reported in the main text.

\subsubsection{Full protection}\label{sec:electric_full}
The temporally averaged dynamics of the electric field under the faulty theory $H=H_0+\lambda H_1+VH_G$ is shown in Fig.~\ref{fig:FullProtection_efield}, at a fixed gauge-breaking error strength $\lambda$ and for various values of the link spin length $S$ and the protection strength $V$. As in the case of the particle density, see Eq.~\eqref{eq:density}, the dynamics agrees very well with that under the adjusted gauge theory $H_\text{adj}=H_0+\lambda\mathcal{P}_0H_1\mathcal{P}_0=H_0$. As the insets in Fig.~\ref{fig:FullProtection_efield} show, the error grows within the analytically derived upper bound of $t^2V_0^3/V$ (details of the derivation can be found in Appendix~\ref{sec:constrained}) at sufficiently large $V$ and for all link spin lengths $S$ considered.

The finite-size behavior of the electric field, displayed in Fig.~\ref{fig:FSSfull_efield}(a,b) for $S=1/2$ and $1$, respectively, shows quick convergence to the thermodynamic limit already at a few matter sites, similarly to the cases of the gauge violation (Fig.~\ref{fig:FSSfull}) and particle density (Fig.~\ref{fig:FSSfull_sigmaz}).

\subsubsection{Linear protection}\label{sec:electric_lin}
\begin{figure*}[htp]
	\centering
	\includegraphics[width=.48\textwidth]{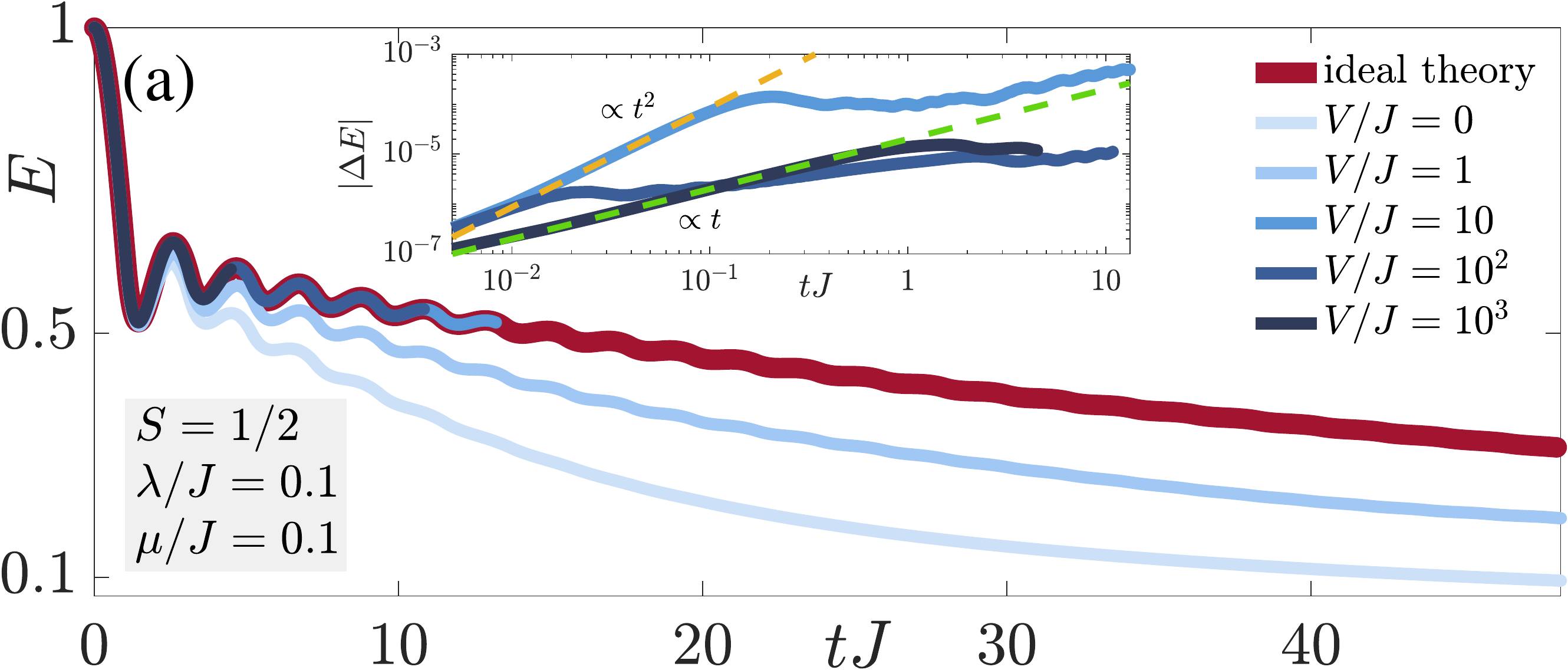}\quad
	\includegraphics[width=.48\textwidth]{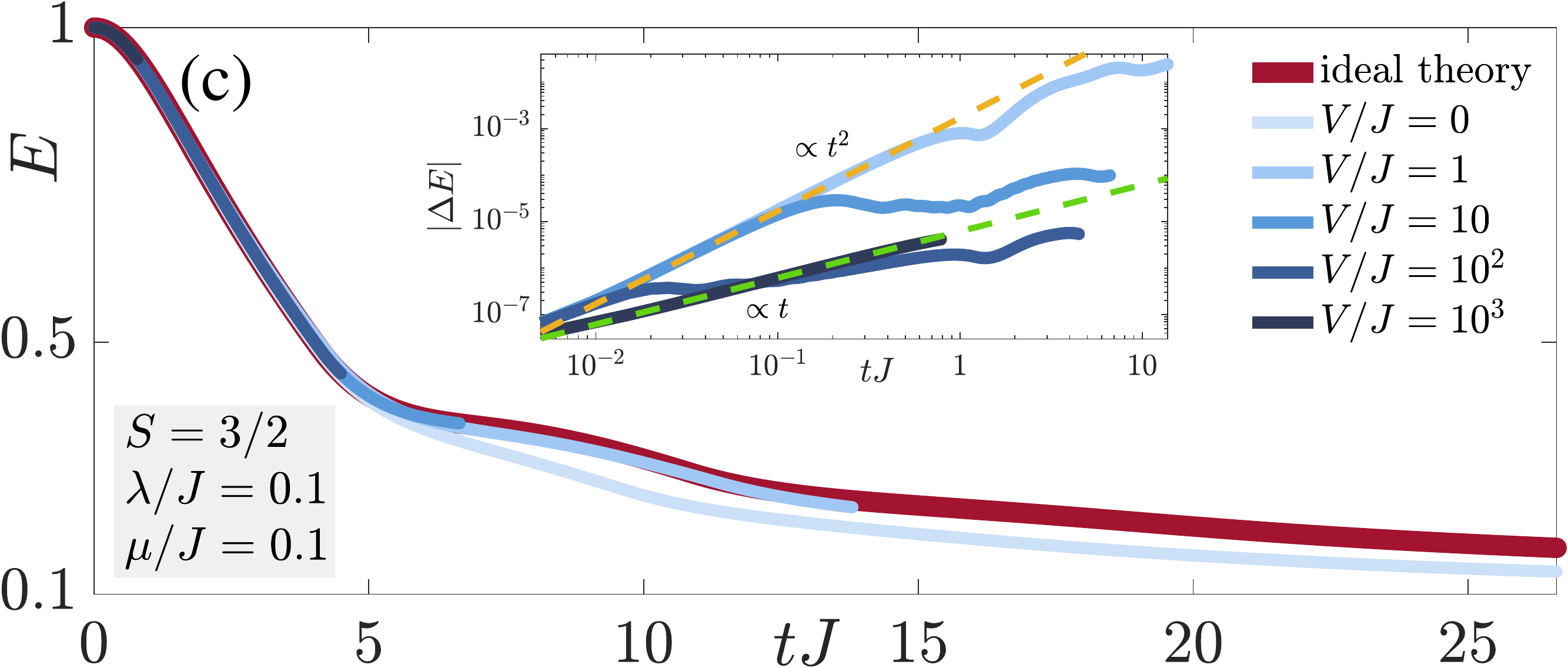} \\
	\includegraphics[width=.48\textwidth]{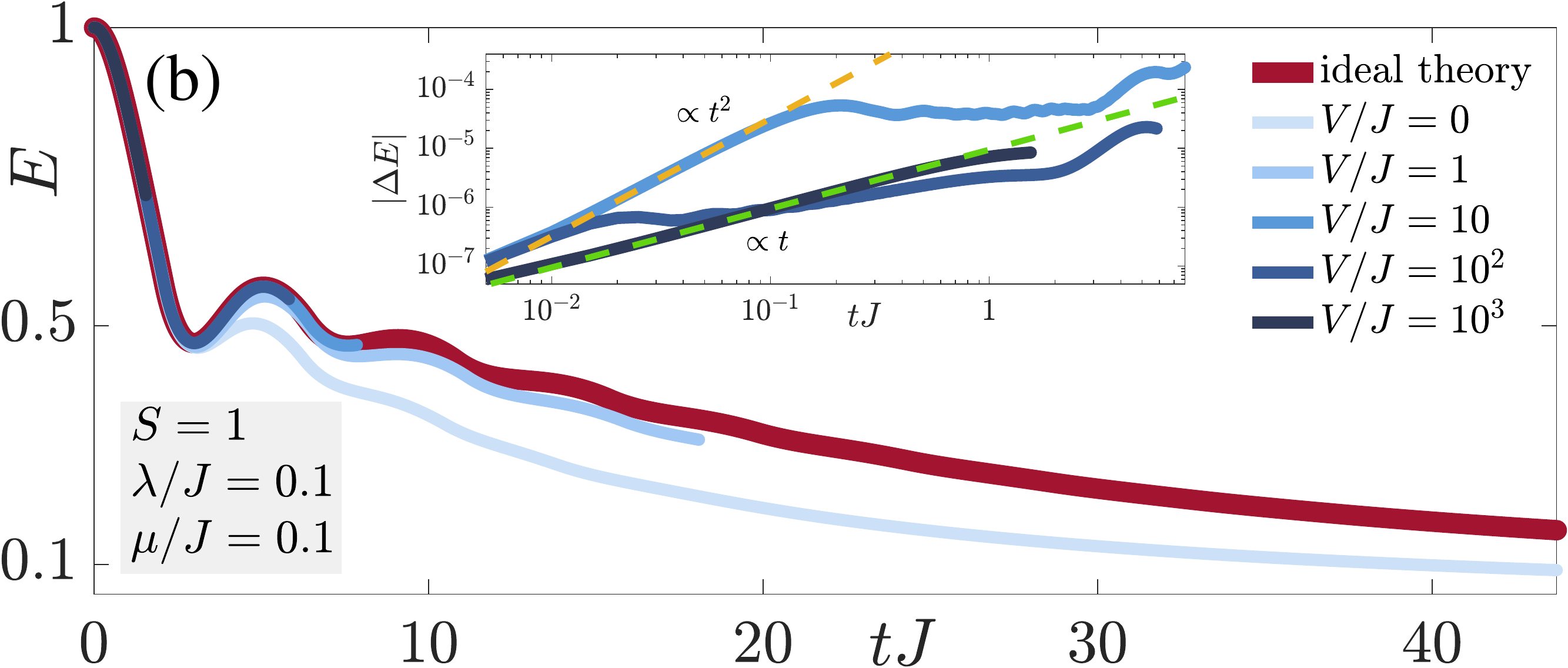}\quad
	\includegraphics[width=.48\textwidth]{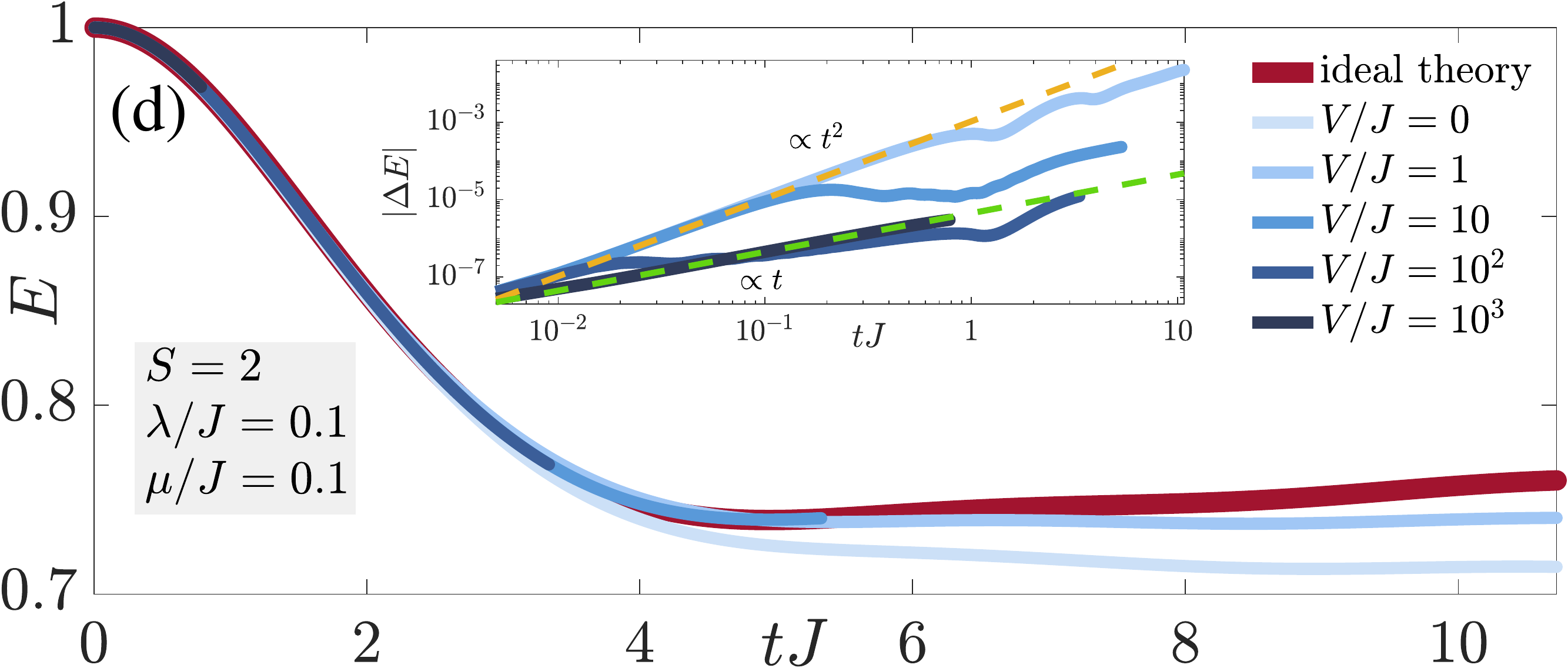}
	\caption{(Color online). Same as Fig.~\ref{fig:FullProtection_efield} but using a linear protection term $V\tilde{H}_G=V\sum_jc_jG_j$ with the noncompliant sequence $c_j=(-1)^{j+1}$. Same as in the case of the particle density in Fig.~\ref{fig:LinProtection_sigmaz} of the main text, the dynamics of the electric field under the faulty gauge theory $H=H_0+\lambda H_1+V\tilde{H}_G$ reliably approaches the ideal-theory dynamics under $H_0$ (which is also the adjusted gauge theory) for $V\gtrsim10$. The insets show the deviation $\lvert\Delta E\rvert$ of the faulty dynamics from its ideal counterpart, where the error initially grows at $\propto t^2$, but then slower ($\propto t$) at later times, with this transition occurring at the timescale $\propto1/V$ as then the protection term begins to dominate. This is milder than the upper error bound with respect to the adjusted gauge theory (see main text and Appendix~\ref{sec:QZE}).
	}
	\label{fig:LinProtection_efield} 
\end{figure*}

\begin{figure}[htp]
	\centering
	\includegraphics[width=.48\textwidth]{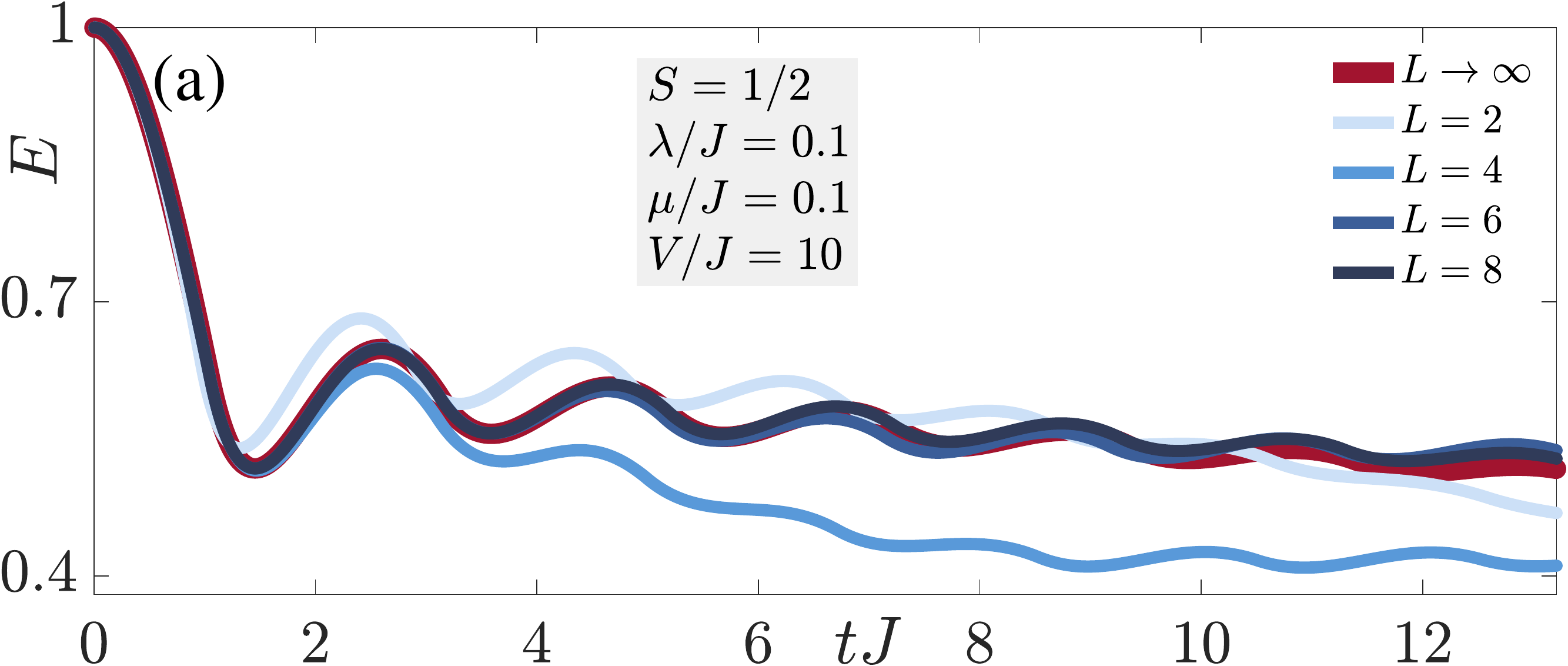}\\
	\includegraphics[width=.48\textwidth]{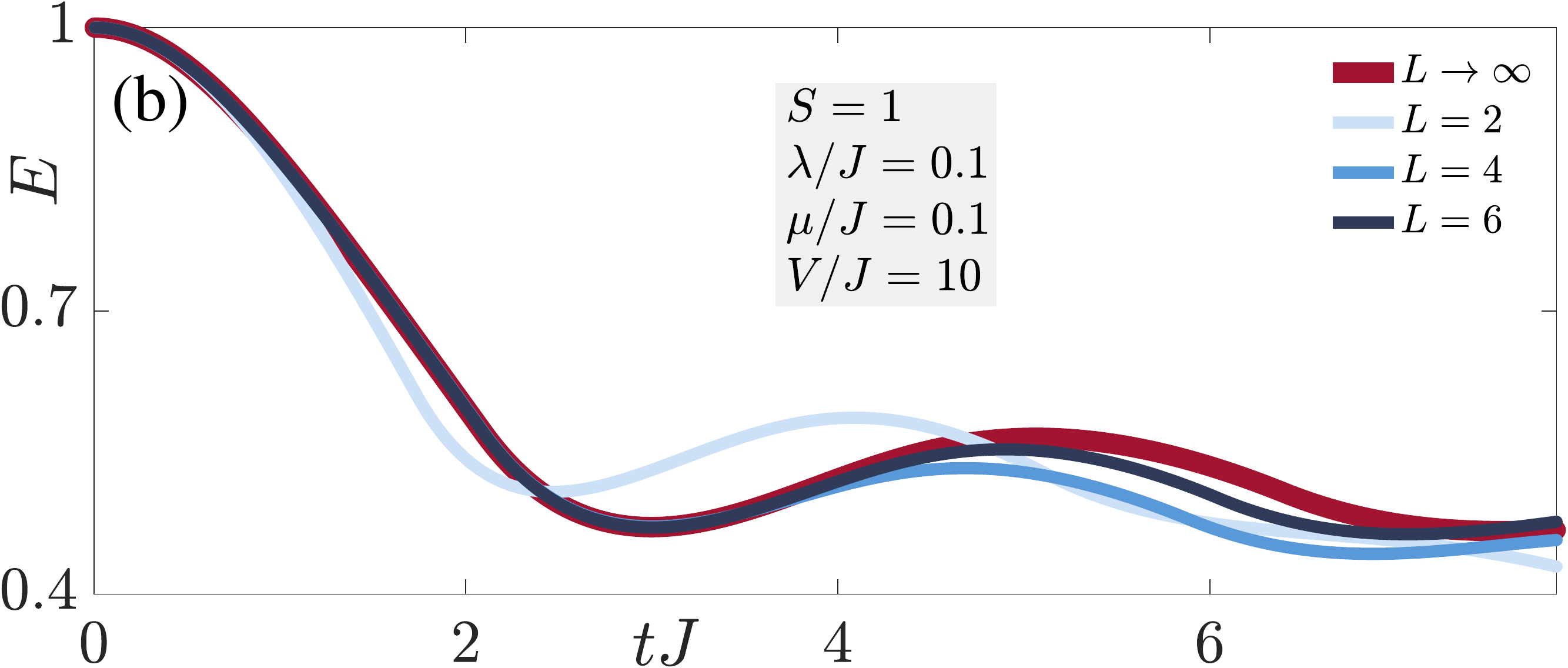}
	\caption{(Color online). Same as Fig.~\ref{fig:FSSfull_efield} but for linear protection with a noncompliant sequence, $V\tilde{H}_G=V\sum_j(-1)^{j+1}G_j$. As in the case of the gauge violation in Fig.~\ref{fig:FSSlin} and particle denstiy in Fig.~\ref{fig:FSSlin_sigmaz} for this protection scheme, the dynamics quickly approaches the thermodynamic limit already at a few matter sites.}
	\label{fig:FSSlin_efield} 
\end{figure}

Similarly to its counterpart for the particle density (Fig.~\ref{fig:LinProtection_sigmaz}), the dynamics of the electric field in the case of linear protection with a noncompliant sequence, $V\tilde{H}_G=V\sum_j(-1)^{j+1}G_j$, shows great agreement with that under the adjusted gauge theory $H_\text{adj}=H_0$ for all accessible times; see Fig.~\ref{fig:LinProtection_efield}. This is particularly surprising because the error upper bound for the adjusted gauge theory in the case of linear protection is $\propto tV_0^2L^2/V$, which becomes infinite in the thermodynamic limit. Nevertheless, our iMPS results indicate that this upper bound is extremely conservative, at least for the scenario considered here. Furthermore, the finite-size behavior of the electric field in this case also shows fast convergence to the thermodynamic limit for all accessible evolution times in iMPS, as exhibited in Fig.~\ref{fig:FSSlin_efield}. This is especially encouraging for ongoing experimental efforts to realize reliable lattice gauge theories, because a single-body protection term such as $V\tilde{H}_G=V\sum_j(-1)^{j+1}G_j$ is straightforward to engineer, e.g., through an optical superlattice in cold-atom experiments.\cite{Yang2020}


\subsection{Temporally nonaveraged gauge violation}\label{sec:raw}
\begin{figure}[htp]
	\centering
	\includegraphics[width=.48\textwidth]{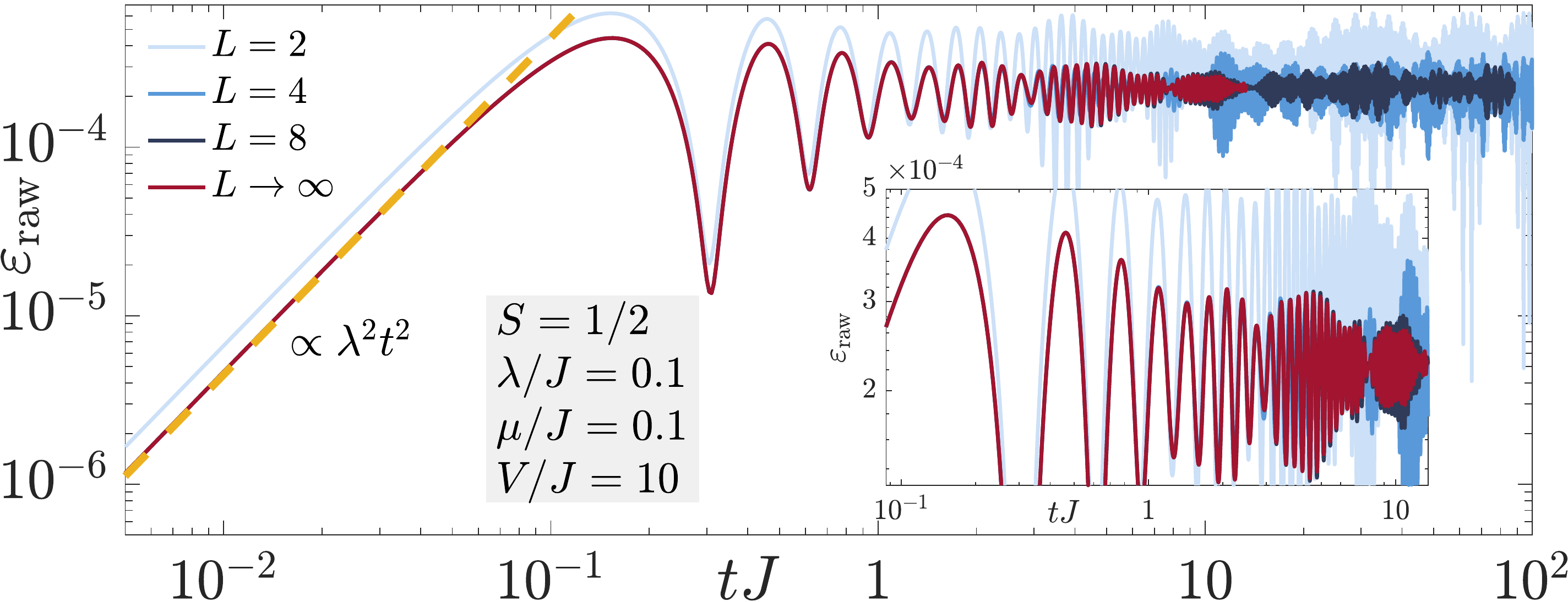}
	\caption{(Color online). Same as Fig.~\ref{fig:FSSlin}(a), but instead of showing the temporally averaged gauge violation $\varepsilon(t)$ given in Eq.~\eqref{eq:violation}, we plot the raw gauge violation $\varepsilon_\text{raw}(t)=\sum_j\bra{\psi_0}e^{iHt}G_j^2e^{-iHt}\ket{\psi_0}/L$. At small system sizes, the fluctuations in the signal decrease and converge fast to the thermodynamic limit.}
	\label{fig:FSSlin_raw}
\end{figure}

For completeness, here we present the behavior of the gauge violation without temporal averaging, given by
\begin{align}
	\varepsilon_\text{raw}(t)=\frac{1}{L}\sum_j\bra{\psi_0}e^{iHt}G_j^2e^{-iHt}\ket{\psi_0}.
\end{align}
The results are presented in Fig.~\ref{fig:FSSlin_raw} for the case of a quench on the initial state $\ket{\psi_0}$ (see Fig.~\ref{fig:illustration}) by the faulty Hamiltonian $H=H_0+\lambda H_1+V\sum_j(-1)^{j+1}G_j$ at different system sizes $2L$ ($L$ matter sites and $L$ links given that we use periodic boundary conditions in our ED calculations). The link spin length is $S=1/2$, while the error and protection strengths are set to $\lambda=0.1J$ and $V=10J$, respectively. As can be seen, temporal fluctuations of $\varepsilon_\text{raw}(t)$ decrease with system size at small values of $L$, but already at about $L=4$ matter sites the fluctuations behave very similarly to the thermodynamic limit. Results at finite values of $L$ are computed in ED, while those in the thermodynamic limit are calculated through iMPS. We have checked that the behavior seen in Fig.~\ref{fig:FSSlin_raw} also applies to other cases, including full protection, different initial states, and other various values of the microscopic parameters.

\begin{figure}[htp]
	\centering
	\includegraphics[width=.48\textwidth]{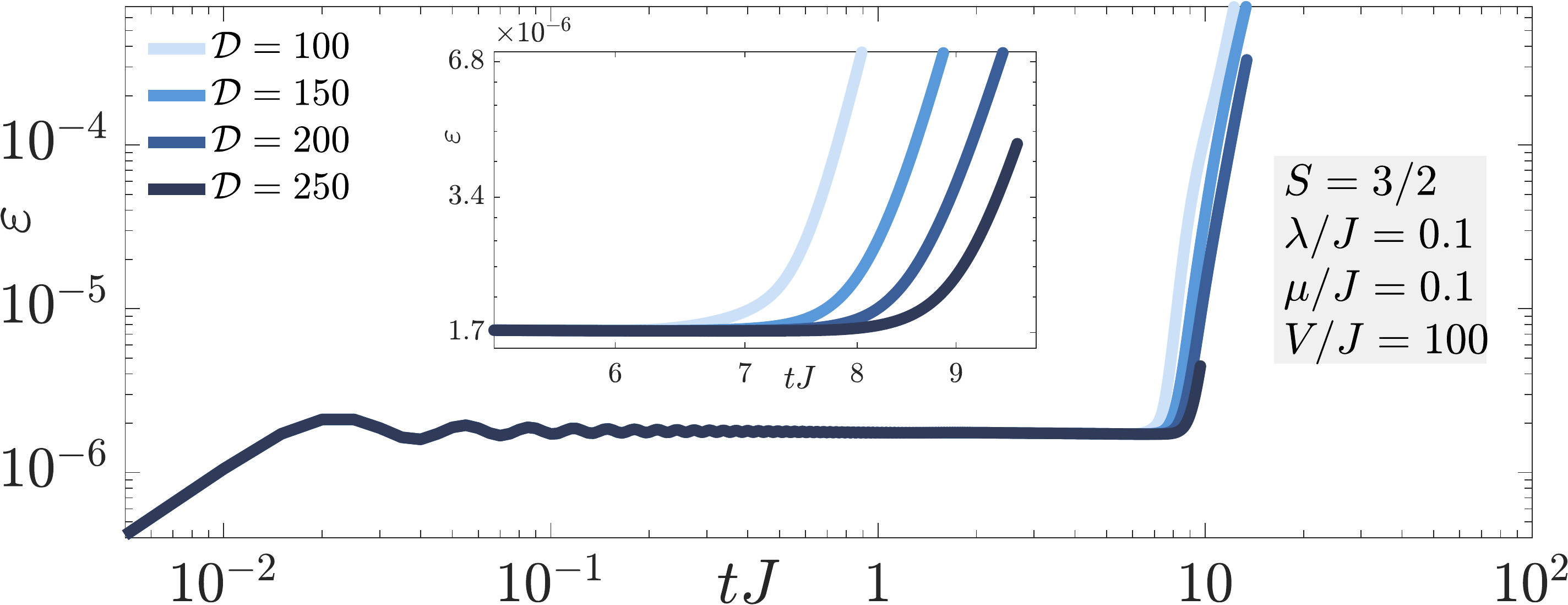}
	\caption{(Color online). Behavior of the gauge violation, Eq.~\eqref{eq:violation}, with respect to iMPS bond dimension at fixed time-step $\Delta t=0.005/J$, in the case of a quench by the faulty Hamiltonian with linear protection $V\tilde{H}_G=V\sum_j(-1)^jG_j$, at error strength $\lambda=0.1J$, protection strength $V=100J$, and for link spin length $S=3/2$ (see legend). As we increase the bond dimension, the plateau $\propto\lambda^2/V^2$ persists longer in evolution time, until the result diverges (see inset). The conclusions of this plot remain the same under full protection, and also for other initial states and different values of the microscopic parameters.}
	\label{fig:bond}
\end{figure}

\subsection{Convergence with bond dimension in iMPS}\label{sec:bond}
In the iMPS results we have shown thus far, we only included data up to evolution times where convergence with respect to time-step and bond dimension is achieved. For our most stringent calculations, this occurs for a time-step $\Delta t=0.005/J$ and bond dimension $\mathcal{D}=200$. However, it is interesting to look at the behavior of the gauge violation relative to $\mathcal{D}$. For this purpose, we again quench the initial state $\ket{\psi_0}$ (see Fig.~\ref{fig:illustration}) by the faulty Hamiltonian $H=H_0+\lambda H_1+V\sum_j(-1)^{j+1}G_j$. The link spin length is $S=3/2$, while the error and protection strengths are set to $\lambda=0.1J$ and $V=100J$, respectively. The time-step is fixed to $\Delta t=0.005/J$. We show in Fig.~\ref{fig:bond} the ensuing gauge violation, Eq.~\eqref{eq:violation}, at several values of the bond dimension $\mathcal{D}$ (see legend). After reaching the plateau $\propto\lambda^2/V^2$, we find that as the bond dimension is increased, the plateau persists to longer evolution times before the result diverges and can no longer be trusted. This behavior is not restricted to the case considered in Fig.~\ref{fig:bond}, but holds also when full protection is used, or the initial state or microscopic parameters are different.

\section{Abanin-De Roeck-Ho-Huveneers (ARHH) method}\label{sec:Abanin}
For the sake of completeness and self-containment, here we briefly review the gauge protection for the full and compliant linear protection terms, introduced in Ref.~\onlinecite{Halimeh2020e} and based on the ARHH framework.\cite{abanin2017rigorous}

For convenience, we define $H_{\rm bare} = H_0 + \lambda H_1$. We also define $\mathcal{P}_n$ as the projection operator onto the eigenstates of ${H}_G$ with eigenvalue $n$. The Hamiltonian $H_{\rm bare}$ can now be decomposed into two parts: $H_{\rm diag}=\sum_n \mathcal{P}_nH_{\rm bare}\mathcal{P}_n = H_0+\lambda \sum_n \mathcal{P}_n H_1\mathcal{P}_n$, and $H_{\rm ndiag}=H-H_{\rm diag}-VH_G$, where $H=H_0+\lambda H_1+VH_G$.

The full protection term $V{H}_G= V\sum_j G_j^2$ protects the gauge-invariant dynamics in the faulty gauge theory $H$ for sufficiently large $V$.\cite{Halimeh2020a,Halimeh2020e,abanin2017rigorous} In the $\mathrm{U}(1)$ gauge theory given in Eq.~\eqref{eq:H0}, the spectrum of ${H}_G$ is composed of integers, based on the form of the corresponding generators $G_j$ of Gauss's law; cf.~Eq.~\eqref{eq:Gj}. Specifically, the ground-state manifold with $n=0$ of ${H}_G$ is the physical sector we want to protect. The projector onto the ground-state manifold is denoted as $\mathcal{P}_0$.

The ARHH method is based on the norm of potentials, which permits a formulation that is volume-independent and applicable in the thermodynamic limit. Define $\Lambda$ as a finite subset of the lattice $\mathbb{Z}^d$, in $d$-dimensional space. The algebra of bounded operators acting on the total Hilbert space $\mathcal{H}_\Lambda$ equipped with the standard operator norm is denoted as $\mathcal{B}_\Lambda$. The subalgebra is defined as $\mathcal{B}_A \subset \mathcal{B}_\Lambda$ of operators of the form $O_A\otimes\mathcal{I}_{\Lambda \backslash A}$ with $A\subset \Lambda$. Generally, any local operator $X$ can be decomposed (in a nonunique way) as $X = \sum_{A\in \mathcal{P}_c(\Lambda)} X_A$ where $X_A \in \mathcal{B}_A$ and $\mathcal{P}_c(\Lambda)$ is the set of finite, connected (by adjacency) subsets of $\Lambda$. We call the collection of $X_A$ a `potential'. The family of norms on potentials with a rate parameter $\kappa \ge 0$, which gives operators with different spatial support different weights, is defined as 
\begin{equation}
\lvert\lvert X\rvert\rvert_\kappa \coloneqq \sup\limits_{x\in\Lambda} \sum_{A\in \mathcal{P}_c(\Lambda):A\ni x} e^{\kappa \lvert A\rvert} \lvert\lvert X_A\rvert\rvert.
\end{equation}
The supremum in this definition picks the lattice site $x$ with the largest sum of weighted norms of the operators $X_A$ that have support on $x$.

With the above definitions, the following can be derived. Given the full protection term $H_G$ with integer spectrum, assume there exists a $\kappa_0$ and relevant energy scale
\begin{align}\label{eq:V0}
V_0 \coloneqq \frac{54\pi}{\kappa_0^2}(\lvert\lvert H_{\rm diag}\rvert\rvert_{\kappa_0}+2\lvert\lvert H_{\rm ndiag}\rvert\rvert_{\kappa_0}).
\end{align} 
Then, if $V$ satisfies the conditions
\begin{subequations}
\begin{align}
\label{eq:Vge}
V &\ge \frac{9\pi\lvert\lvert H_{\rm ndiag}\rvert\rvert_{\kappa_0}}{\kappa_0},\\\label{eq:nstar}
n_* &\coloneqq \bigg\lfloor \frac{V}{V_0(1+\ln V-\ln V_0)^3}\bigg\rfloor - 2 \ge 1,
\end{align}
\end{subequations}
there exists a quasilocal unitary operator $Y$ satisfying
\begin{align}\nonumber
YHY^\dag&=VH_G + H^\prime \\
&= VH_G + H^\prime_{\rm diag} + H^\prime_{\rm ndiag},
\end{align}
where 
\begin{subequations}
\begin{align}
H^\prime &= YHY^\dag - VH_G,\\
H^\prime_{\rm diag}&=\sum_n \mathcal{P}_n H^\prime\mathcal{P}_n,\\ H^\prime_{\rm ndiag}&=H^\prime-H^\prime_{\rm diag},\\
\lvert\lvert H^\prime_{\rm diag}-H_{\rm diag}\rvert\rvert_{\kappa_{n_*}}&\le CV_0/V, \\
\lvert\lvert H^\prime_{\rm ndiag}\rvert\rvert_{\kappa_{n_*}}&\le (2/3)^{n_*}||H_{\rm ndiag}||_{\kappa_0},\\
\kappa_{n_*}&\coloneqq\frac{\kappa_0}{1+\log(1+n_*)},
\end{align}
\end{subequations}
with $C$ a constant. 

For arbitrary local operator $O$ and up to an exponentially large time $t$ on the scale $e^{kn_*}/V_0$, we have 
\begin{align}
&\big\lvert\big\lvert U(t)^\dag OU(t)-e^{it(VH_G+H^\prime_{\rm diag})}Oe^{-it(VH_G+H^\prime_{\rm diag})}\big\rvert\big\rvert\nonumber
\\ 
&\le \frac{K(O)}{V},
\end{align}
where $U(t)=e^{-iHt}$ is the time-evolution operator, $0<k<(d+1)^{-1}\ln{(3/2)}$, and $K(O)$ is $V$- and volume-independent but model parameter-dependent term.

In the operator norm sense, the dynamics of all local observables as generated by the effective Hamiltonian $H^\prime_{\rm diag}$ is perturbatively close (in $V_0/V$) to $H_{\rm diag}$. As discussed in Ref.~\onlinecite{Halimeh2020e}, we cannot expect $H^\prime_{\rm diag}$ to be a gauge-invariant Hamiltonian. However,
if we prepare the initial state in the physical target space and focus on the dynamics of expectation values, the above operator norm bound can be translated to
\begin{align}
\big\lvert\langle U^\dag(t) OU(t)-e^{iH^\prime_{\rm eff}t}Oe^{-iH^\prime_{\rm eff}t}\rangle\big\rvert\le \frac{K(O)}{V},
\end{align}
where $H^\prime_{\rm eff}= \mathcal{P}_0 H^\prime \mathcal{P}_0$. In contrast to $H^\prime_{\rm diag}$, $H^\prime_{\rm eff}$ is a gauge-invariant Hamiltonian. Up to now, we obtain an effective renormalized gauge Hamiltonian $H^\prime_{\rm eff}= \mathcal{P}_0 H^\prime \mathcal{P}_0$, which governs all the dynamics of expectation values of local observables $O$ with an error bounded by $K(O)/V$ up to a timescale $e^{kn_\star}/V_0$. This is what we refer to in the main text as the \textit{renormalized timescale}, which can be written as $\tau_\text{ren}\propto\exp(V/V_0)/V_0$ when $V$ is sufficiently large. It is worth mentioning that this timescale also exists in the case of linear protection with a \textit{compliant} sequence $c_j$ (for derivational details see Ref.~\onlinecite{Halimeh2020e}). However, in that case $V$ has to be increased with system size $L$ in order to achieve a given level of gauge-invariance reliability. 
This is because the larger $L$ is, the smaller is the spacing in the normalized rational coefficients $c_j$ of the compliant sequence. As such, even though the renormalized timescale itself is not explicitly volume-dependent, the reliability is not in the case of linear protection with a \textit{compliant} sequence. This is not a problem in the case of full protection, and therefore the reliability is volume-independent.

Let us now look at how $V$ scales with the link spin length $S$. The exact form of $H^\prime_{\rm eff}$ is in general difficult to obtain. Nevertheless, this framework permits us to make general scaling statements. In particular, the required protection strength is volume-independent and determined by Eqs.~\eqref{eq:Vge} and~\eqref{eq:nstar}. For a perturbative error term $\lambda H_1$ and sufficiently large $S$, Eq.~\eqref{eq:nstar} determines the minimal protection strength as $V_\mathrm{min}\sim V_0$. In this case, $V_0$ roughly scales as the maximum of the norm of the local terms of $H_{\rm bare}$, which in our model is $g^2a(s^z_{j,j+1})^2/2 \sim S^2$. Therefore, in the worst case the minimal protection strength $V_\mathrm{min}$ scales as $S^2$ when $S$ is sufficiently large.

\section{Constrained quantum dynamics}\label{sec:constrained}
As mentioned in Appendix~\ref{sec:Abanin} treating the ARHH method, the dynamics of local observables can be protected up to an exponential timescale. 
However, the fact that the renormalized effective Hamiltonian is in general unknown makes that method somewhat inconvenient for practical applications to concrete systems, particularly when knowledge of an emergent theory is desired to quantify a controlled error. 
Using a simpler framework also in the context of full protection, we can additionally prove that the dynamics is generated by an \textit{adjusted} gauge theory described by the Hamiltonian $H_\text{adj}=H_0 + \lambda \mathcal{P}_0H_1\mathcal{P}_0$ up to a timescale fractional in the protection strength $V$ and is still volume-independent and scales as $S^2$ when $S$ is sufficiently large. For the physical error given by Eq.~\eqref{eq:error}, this adjusted gauge theory Hamiltonian is exactly $H_0$ since $\mathcal{P}_0H_1\mathcal{P}_0=0$ in this case. This section adapts recent work treating a universal bound on the constrained quantum dynamics.\cite{gong2020error,gong2020universal}

The ingredients of the derivations of this bound are: 
\begin{enumerate}
	\item The Hamiltonian $H$ and observable $O$ are local in the sense of the potentials introduced in Appendix~\ref{sec:Abanin}, which is identical to the locality requirement of the ARHH method.
	\item All terms in the decomposition of $H_{\rm pro} = \sum_{A\in \mathcal{P}_c(\Lambda)} H_{{\rm pro}, A}$ commute with one other, where the collection of $H_{{\rm pro}, A}$ is a potential.
	\item The ground-state manifold of $H_{\rm pro}$ is frustration-free, i.e., the ground state also minimizes the local energies $H_{{\rm pro}, A}$ everywhere for all $A$.
	\item There exists a sufficiently large energy gap $\Delta_0$ between the ground state and excited states of $H_{\rm pro}$. By sufficiently large, we mean here that $V$ scales at least $\sim \lvert\lvert H_{\rm bare}\rvert\rvert_{\kappa=0}$.
\end{enumerate}
If all these four conditions are satisfied, the following bound holds:
\begin{widetext}
\begin{align}
	\big\lvert\big\lvert\mathcal{P}_0\big[U(t)^\dag OU(t)-e^{it(H_{\rm pro}+\mathcal{P}_0H_{\rm bare}\mathcal{P}_0)}Oe^{-it(H_{\rm pro}+\mathcal{P}_0H_{\rm bare}\mathcal{P}_0)}\big]\mathcal{P}_0\big\rvert\big\rvert\le \frac{\lvert\lvert H_{\rm bare}\rvert\rvert_{\kappa=0}}{\Delta_0}p(t),
\end{align}
\end{widetext}
where $p(t)$ is a polynomial in $||H_{\rm bare}||_{\kappa=0}t$ with degree $d + 1$ and (at most) order-one coefficients.

It is easy to check that the model $H=H_0+\lambda H_1+VH_G$ in this work satisfies all these necessary conditions. The decomposition of $H_{\rm pro}=VH_G$ can be chosen as $H_{{\rm pro},A} = VG_j^2$ and the energy gap of $H_{\rm pro}$ is $V$. To apply the bound to $H$, we immediately obtain that the dynamics of the arbitrary local observable $O$ is generated by the adjusted Hamiltonian $H_\text{adj}=H_0 + \lambda \mathcal{P}_0H_1\mathcal{P}_0$ with an error bounded by $\lvert\lvert H_0 + \lambda H_1\rvert\rvert_{\kappa=0}/V$ up to a timescale fractional in $V$, which is volume-independent,
\begin{align}
	&\big\lvert\langle U^\dag(t) OU(t)-e^{iH_{\rm adj}t}Oe^{-iH_{\rm adj}t}\rangle\big\rvert\nonumber
	\\ 
	&\le \frac{||H_{\rm bare}||_{\kappa = 0}}{V}p(t) \sim \frac{t^2V_0^3}{V},
\end{align}
In the above expression, we have already used $V_0$ instead of $||H_{\rm bare}||_{\kappa = 0}$ since they are roughly of the same scale in magnitude. In the large $S$ limit, as discussed in the Appendix~\ref{sec:Abanin}, $||H_0 + \lambda H_1||_{\kappa=0}$ scales as $S^2$, which means the protection strength is volume-independent and scales as $S^2$ in the worst case.

\section{Quantum Zeno effect}\label{sec:QZE}
In Ref.~\onlinecite{Halimeh2020e}, it has been shown that linear protection with the noncompliant sequence of coefficents $c_j=(-1)^{j+1}$ can protect the evolution operator $\exp(-iHt)$ of the spin-$1/2$ $\mathrm{U}(1)$  gauge-theory dynamics because of the quantum Zeno effect,\cite{facchi2002quantum,facchi2004unification,facchi2009quantum,burgarth2019generalized}
\begin{align}
	\big\lvert\big\lvert e^{-iHt} - e^{-i(H_{\rm adj} + H_{\rm pro} )t}\big\rvert\big\rvert \le \frac{q(t)}{V},
\end{align}
where $q(t)$ is a polynomial in $t(V_{0,\rm qze}L)^2$ of degree $1$ and a coefficient of order $1$, and $V_{0,{\rm qze}}\sim V_0$ is the relevant energy scale for the quantum Zeno effect. Therefore, the deviation of observables $O$ is given by
\begin{widetext}
\begin{align}
	&\big\lvert\langle U^\dag(t) OU(t)-e^{iH_{\rm adj}t}Oe^{-iH_{\rm adj}t}\rangle\big\rvert  \le  \big\lvert\big\lvert\langle U^\dag(t) OU(t)-e^{iH_{\rm adj}t}Oe^{-iH_{\rm adj}t}\rangle\big\rvert\big\rvert \nonumber
	\\ 
	&=\big\lvert\big\lvert \big[U^\dag(t) - e^{iH_{\rm adj}t}\big] OU(t) + \big[U^\dag(t) - e^{iH_{\rm adj}t} \big]Oe^{-iH_{\rm adj}t} + U^\dag(t) O\big[U(t)- e^{-iH_{\rm adj}t}\big]+  e^{iH_{\rm adj}t} O \big[U^\dag(t) - e^{-iH_{\rm adj}t}\big]\big\rvert\big\rvert \nonumber
	\\
	&\le 4\frac{q(t)}{V} \sim \frac{tV_0^2L^2}{V}.
\end{align}
\end{widetext}
Hence, the gauge theory is reliable up to a timescale $t\propto V/(V_{0,{\rm qze}}L)^2\sim V/(V_{0}L)^2$, with a controlled violation of $\mathcal{O}(V_{0,{\rm qze}}^2L^2/V)\sim \mathcal{O}(V_{0}^2L^2/V)$.

\bibliography{Reliability_iMPS}
\end{document}